\documentstyle[12pt,aaspp4,flushrt]{article}

\newcommand\about     {\hbox{$\sim$}}

\newcommand\RRc       {candidate RR Lyrae stars}
\newcommand\othername {\hbox{$\dots$}}   

\begin{document}

\title{         Candidate RR Lyrae Stars Found in 
          Sloan Digital Sky Survey Commissioning Data $^1$   }

\author{
\v{Z}eljko Ivezi\'{c}\altaffilmark{\ref{Princeton}},
Josh Goldston\altaffilmark{\ref{Princeton}},
Kristian Finlator\altaffilmark{\ref{Princeton}},
Gillian R. Knapp\altaffilmark{\ref{Princeton}}, 
Brian Yanny\altaffilmark{\ref{Fermilab}},
Timothy A. McKay\altaffilmark{\ref{Michigan}},
Susan Amrose\altaffilmark{\ref{Michigan}},
Kevin Krisciunas\altaffilmark{\ref{Washington}}, 
Beth Willman\altaffilmark{\ref{Washington}}, 
Scott Anderson\altaffilmark{\ref{Washington}}, 
Chris Schaber\altaffilmark{\ref{Washington}},
Dawn Erb\altaffilmark{\ref{Washington}}, 
Chelsea Logan\altaffilmark{\ref{Washington}},
Chris Stubbs\altaffilmark{\ref{Washington}},
Bing Chen\altaffilmark{\ref{JHU}},
Eric Neilsen\altaffilmark{\ref{JHU}},
Alan Uomoto\altaffilmark{\ref{JHU}}, 
Jeffrey R. Pier\altaffilmark{\ref{Flagstaff}},
Xiaohui Fan\altaffilmark{\ref{Princeton}},
James E. Gunn\altaffilmark{\ref{Princeton}},
Robert H. Lupton\altaffilmark{\ref{Princeton}},
Constance M. Rockosi\altaffilmark{\ref{Chicago}},
David Schlegel\altaffilmark{\ref{Princeton}},
Michael A. Strauss\altaffilmark{\ref{Princeton}},
James Annis\altaffilmark{\ref{Fermilab}},
Jon Brinkmann\altaffilmark{\ref{APO}},
Istv\'an Csabai\altaffilmark{\ref{JHU},\ref{Eotvos}},
Mamoru Doi\altaffilmark{\ref{UTokyo}},
Masataka Fukugita\altaffilmark{\ref{CosmicRay},\ref{IAS}},
Gregory S. Hennessy\altaffilmark{\ref{USNO}},
Robert B. Hindsley\altaffilmark{\ref{RSD}},
Bruce Margon\altaffilmark{\ref{Washington}}, 
Jeffrey A. Munn\altaffilmark{\ref{Flagstaff}},
Heidi Jo Newberg\altaffilmark{\ref{RPI}},
Donald P. Schneider\altaffilmark{\ref{PennState}},
J. Allyn Smith\altaffilmark{\ref{Michigan}},
Gyula P. Szokoly\altaffilmark{\ref{Potsdam}},
Aniruddha R. Thakar\altaffilmark{\ref{JHU}},
Michael S. Vogeley\altaffilmark{\ref{Princeton}},
Patrick Waddell\altaffilmark{\ref{Washington}},
Naoki Yasuda\altaffilmark{\ref{NAOJapan}}, and
Donald G. York\altaffilmark{\ref{Chicago}}
for the SDSS Collaboration
}

\altaffiltext{1}{Based on observations obtained with the
Sloan Digital Sky Survey.}
\newcounter{address}
\setcounter{address}{2}
\altaffiltext{\theaddress}{Princeton University Observatory, Princeton, NJ 08544
\label{Princeton}}
\addtocounter{address}{1}
\altaffiltext{\theaddress}{Fermi National Accelerator Laboratory, P.O. Box 500,
Batavia, IL 60510
\label{Fermilab}}
\addtocounter{address}{1}
\altaffiltext{\theaddress}{University of Michigan, Department of Physics,
500 East University, Ann Arbor, MI 48109
\label{Michigan}}
\addtocounter{address}{1}
\altaffiltext{\theaddress}{University of Washington, Department of Astronomy,
Box 351580, Seattle, WA 98195
\label{Washington}}
\addtocounter{address}{1}
\altaffiltext{\theaddress}{U.S. Naval Observatory, Flagstaff Station,
P.O. Box 1149,
Flagstaff, AZ  86002-1149
\label{Flagstaff}}
\addtocounter{address}{1}
\altaffiltext{\theaddress}{University of Chicago, Astronomy \& Astrophysics
Center, 5640 S. Ellis Ave., Chicago, IL 60637
\label{Chicago}}
\addtocounter{address}{1}
\altaffiltext{\theaddress}{Apache Point Observatory, P.O. Box 59,
Sunspot, NM 88349-0059
\label{APO}}
\addtocounter{address}{1}
\altaffiltext{\theaddress}{Department of Physics and Astronomy, 
The Johns Hopkins University, 3701 San Martin Drive, Baltimore, MD 21218
\label{JHU}}
\addtocounter{address}{1}
\altaffiltext{\theaddress}{Department of Physics of Complex Systems,
E\"otv\"os University, P\'azm\'any P\'eter s\'et\'any 1/A, Budapest, H-1117, Hungary
\label{Eotvos}}
\addtocounter{address}{1}
\altaffiltext{\theaddress}{Department of Astronomy and Research Center for 
the Early Universe, School of Science, University of Tokyo, Hongo, Bunkyo,
Tokyo, 113-0033 Japan
\label{UTokyo}}
\addtocounter{address}{1}
\altaffiltext{\theaddress}{Institute for Cosmic Ray Research, University of
Tokyo, Midori, Tanashi, Tokyo 188-8502, Japan
\label{CosmicRay}}
\addtocounter{address}{1}
\altaffiltext{\theaddress}{Institute for Advanced Study, Olden Lane,
Princeton, NJ 08540
\label{IAS}}
\addtocounter{address}{1}
\altaffiltext{\theaddress}{U.S. Naval Observatory,
3450 Massachusetts Ave., NW,
Washington, DC  20392-5420
\label{USNO}}
\addtocounter{address}{1}
\altaffiltext{\theaddress}{Remote Sensing Division, Code 7215, Naval Research 
Laboratory, 4555, Overlook Ave. SW, Washington, DC 20375
\label{RSD}}
\addtocounter{address}{1}
\altaffiltext{\theaddress}{Department of Physics, Applied Physics and Astronomy, 
Rensselaer Polytechnic Institute, Troy, NY 12180
\label{RPI}}
\addtocounter{address}{1}
\altaffiltext{\theaddress}{Department of Astronomy and Astrophysics,
The Pennsylvania State University,
University Park, PA 16802
\label{PennState}}
\addtocounter{address}{1}
\altaffiltext{\theaddress}{Astrophysikalisches Institut Potsdam, An der 
Sternwarte 16, D-14482 Potsdam, Germany
\label{Potsdam}}
\addtocounter{address}{1}
\altaffiltext{\theaddress}{National Astronomical Observatory, 2-21-1, Osawa, 
Mitaka, Tokyo 181-8588, Japan
\label{NAOJapan}}

\begin{abstract}     

We present a sample of 148 \RRc\ selected from Sloan Digital Sky
Survey (SDSS) commissioning data for about 100 deg$^2$ of sky surveyed 
twice with $\Delta t$ =1.9946 days. Although the faint magnitude limit of
the SDSS allows us to detect RR Lyrae stars to large galactocentric distances 
($\sim$ 100 kpc, or $r^* \sim$ 21), we find no candidates fainter than 
$r^* \sim$ 20, i.e. further than  $\sim$ 65 kpc from the Galactic center. 
On the assumption that all 148 candidates are indeed RR Lyrae stars (contamination
by other species of variable star is probably less than 10\%), we find that 
their volume density has roughly a power-law dependence on galactocentric 
radius, $R^{-2.7\pm0.2}$, between 10 and 50 kpc, and drops abruptly at 
$R \sim$ 50--60 kpc, possibly indicating a sharp edge to the stellar halo 
as traced by RR Lyrae stars. 

The Galactic distribution of stars in this sample is very inhomogeneous 
and shows a clump of over 70 stars at about 45 kpc from the Galactic center. 
This clump is also detected in the distribution of nonvariable objects with 
RR Lyrae star colors. When sources in the clump are excluded, the best 
power-law fit becomes consistent with the $R^{-3}$ distribution found from 
surveys of bright RR Lyrae stars. These results imply that the halo contains 
clumpy overdensities inhomogeneously distributed within a smooth $R^{-3}$ 
background, with a possible cutoff at $\sim$ 50 kpc.

\end{abstract}


\section{             Introduction              }

The SDSS is a digital photometric and spectroscopic survey which will 
cover one quarter of the Celestial sphere in the north Galactic cap and produce
a smaller area ($\sim$ 225 deg$^2$) but much deeper survey in the southern 
Galactic hemisphere (Gunn \& Weinberg 1995, York {\em et al.} 2000\footnote{see 
also http://www.astro.princeton.edu/PBOOK/welcome.htm}). The flux densities of
detected objects are measured almost simultaneously in five bands ($u'$, $g'$, $r'$, 
$i'$, and $z'$, \cite{F96}) with effective wavelengths of 3540~\AA, 
4760~\AA, 6280~\AA, 7690~\AA, and 9250~\AA, complete to limiting 
(5:1 signal-to-noise) point source magnitudes of 22.3, 23.3, 23.1, 
22.3, and 20.8 in the North Galactic cap\footnote{We refer to the measured 
magnitudes as $u^*, g^*, r^*, i^*,$ and $z^*$ 
because the absolute calibration of the SDSS photometric system is still 
uncertain at the $\sim 0.05^m$ level. The SDSS filters themselves are 
referred to as $u', g', r', i',$ and $z'$. All magnitudes are given on 
the AB$_\nu$ system (Oke \& Gunn, 1983; for additional discussion regarding 
the SDSS photometric system see \cite{F96} and Fan 1999).}. 
The survey sky coverage of 
about $\pi$ steradians will result in photometric measurements to 
the above detection limits for about 10$^8$ stars. The pixel size (0.4 arcsec) 
and optical quality of the telescope are such that the resolution is 
limited by atmospheric seeing. Astrometric positions are accurate to about 
0.1 arcsec for sources brighter than 20.5$^m$, and the morphological and color 
information from the images allows robust star-galaxy separation to 
$\sim$ 21.5$^m$. 

The survey is being done with a dedicated, special-purpose 2.5 meter
telescope (\cite{Siegmund00}). It has a wide well-corrected field (3 deg)
and is equipped with a large mosaic CCD camera and a pair of fiber-fed spectrographs. 
The camera utilizes thirty large-area (2048 x 2048) CCDs (\cite{Gunnetal}, Doi 
{\em et al.} 2000) which take the data in drift-scanning (time-delay-and-integrate, 
or TDI) mode with a total integration time of 54.1 s. The imaging data are 
obtained using the data acquisition system at the Apache Point Observatory 
(Petravick {\em et al.} 1994, Annis {\em et al.} 2000) and recorded on DLT tapes. 
These tapes are shipped to Fermilab by express courier and the data are
automatically reduced through a set of software pipelines
operating in a common computing environment (Kent {\em et al.} 2000). The 
photometric pipeline (\cite{Lupton00}) reduces the imaging data, 
measuring positions, magnitudes, and shape parameters for all detected objects. The 
photometric pipeline uses position calibration information from the
astrometric pipeline (\cite{Pier00}) and photometric calibration data
from the photometric telescope (Smith {\em et al.} 2000, 
Uomoto {\em et al.} 2000), reduced through
the photometric telescope pipeline (Tucker {\em et al.} 2000). Final
calibrations are applied by the final calibration pipeline which allows
refinements in the positional and photometric calibration to be applied
as the survey progresses. The outputs, together with all the observing
and processing information, are loaded into the operational data base 
(Yanny {\em et al.} 2000b), which is the central repository of scientific 
and bookkeeping data used to run the survey.

About 40\% of the sky in the northern survey will be surveyed twice
(because of the scan overlaps), and all of the southern survey dozens 
of times (to search for variable objects and, by stacking the frames, to go 
deeper). Although two observations are normally insufficient to characterize 
a variable  object, the multi-color nature of the photometric data helps 
enormously. Close to 1000 square degrees of sky along the Celestial Equator 
have been observed during the SDSS commissioning phase (e.g. \cite{F99}). 
About 100 deg$^2$ of sky have been observed more than once; we use these data 
to search for variable objects. In this paper we describe the detection and 
analysis of $\sim$ 150 variable objects, probably RR Lyrae stars. 
Detecting RR Lyrae stars is important for Galactic structure studies because:
 
\begin{itemize}
\item
They are believed to be an unbiased tracer of the low-metallicity halo population 
for kinematic studies (\cite{H84}),
\item 
They are nearly standard candles ($\langle M_V \rangle$ = 0.7 $\pm$ 0.1, \cite{L96})
and thus it is straightforward to determine their distance, and
\item 
They are sufficiently bright to be detected at large distances ($\sim$ 100
kpc for $r'$ $\sim$ 21$^m$) and are thus especially well-suited 
for studies of the outer halo (\cite{S84}). 
\end{itemize}

For a comprehensive review of RR Lyrae stars we refer the reader to 
Smith (1996).

Wetterer \& McGraw (1996) recently compiled data from several 
available deep searches for RR Lyrae stars in the outer Galactic halo.
They pointed out that there are only nine RR Lyrae stars discovered at
galactocentric distances larger than 30 kpc, and only a few more have 
been found since then (e.g. Margon \& Deutsch 1999, and references 
therein). This paper presents a sample of candidate RR Lyrae stars with 
81 stars estimated to be at such distances. The following section describes 
the selection procedure used in the search for \RRc\ and the resulting sample. 
We analyze the Galactic distribution of selected stars in \S3, and compare the
sample to two other surveys of variable stars in \S4.

\section         {        Selection Procedure        }

We utilize imaging data from four runs (77, 745, 752, and 756) obtained 
during the SDSS commissioning phase. The data were obtained in six 
parallel scanlines\footnote{See also 
http://www.astro.princeton.edu/PBOOK/strategy/strategy.htm}, 
each 13.5 arcmin wide, along the Celestial Equator 
($-1.2687^\circ <$ Dec $< 1.2676^\circ$). The seeing in all runs was variable 
between 1 and 2 arcsec (FWHM) with the median value typically 1.5 arcsec. 
Runs 745 and 756, which are used as the main data set, were taken 1.9946 
days apart on March 20 and 22, 1999. Their overlap extends from RA 
= 10$^h$ 42$^m$ ($l=248.6^\circ$, $b=48.7^\circ$) to RA = 15$^h$ 46$^m$ 
($l$=7.3$^\circ$, $b$=40.2$^\circ$) and covers 97.5 deg$^2$ of sky (all 
coordinates are given as J2000). 

As auxiliary control data sets we use overlaps between runs 77 and 745 
(33.7 deg$^2$), 77 and 756 (20.5 deg$^2$), and 752 and 756 (11.1 deg$^2$).
The RA ranges for these overlaps are summarized in Table 1. Run 77 was 
obtained on June 27, 1998 and provides a $\sim 9$ month (266 and 268 days) 
baseline when compared to runs 745 and 756. Run 752 was obtained on March 21, 
1999 and provides a baseline of 0.9976 days when combined with run 756. 
The overlap between runs 752 and 756 perpendicular to the scan direction 
(i.e. declination overlap) is only 12.4\% of the scanline width (these are two 
six-column strips which are later interleaved to make a filled stripe of 
imaging data), while the overlaps between runs 77-745, 77-756 and 745-756 
are 67.7\%, 63.7\% and 95.9\% of the scanline width, respectively (scanline 
width = 6 $\times$ 0.2253 deg = 1.3517 deg).

RR Lyrae stars have the colors of A and F stars (Preston 1959; for 
SDSS colors see \cite{F96}, \cite{KMS98}; note also Figure 2 below)
and thus could be selected by appropriately constraining all four 
SDSS colors, and then searching for variability. 
However, this would produce complicated selection effects because not all 
five bands have the same sensitivity. In order to avoid such effects we 
follow a two-step procedure; the first step uses only 
data from the most sensitive $g'$ and $r'$ bands, and only in the second 
step do we introduce color cuts based on data in all five bands. Throughout this 
work we use the ``point spread function'' (PSF) magnitudes (measured by 
fitting a PSF model of a double Gaussian) as computed by the photometric 
pipeline (``photo'', version v5\_1, for details see \cite{Lupton00}).

Since RR Lyrae stars are bluer in $g^* - r^*$ than most other stars, we 
started our search by selecting the 90569 unresolved and unsaturated sources 
with $-0.1 < g^* - r^* < 0.4$ from $\sim$ 930,000 stars in the overlap of
runs 745 and 756, and then required that candidate variable sources satisfied
the following conditions:

\begin{enumerate}
\item
The difference between the magnitudes in the two runs in both $g'$ and $r'$
bands is at least 0.15$^m$.
\item
The difference between the magnitudes in the two runs in both $g'$ and $r'$
bands is at least 5$\sigma$. Here the errors are taken as estimated by 
the photometric pipeline and do not include systematic calibration errors. 
However, because of the above requirement (variability of at least 0.15$^m$), 
this condition becomes relevant only at the faint end where the errors
are dominated by photon statistics (see Figure 1 below).
\item
Candidates are brighter in $r^*$ when they are bluer in $g^* - r^*$ (since
RR Lyrae stars are pulsating variable stars). This is equivalent to the 
condition that the difference between the magnitudes in $g'$ band is larger 
than the difference between the magnitudes in $r'$ band. This condition is
implemented without accounting for photometric errors.
\end{enumerate}

These selection criteria yield 186 candidates. In Figure 1a, the large dots
show the observed change in the $r^*$ magnitude for the selected candidates 
plotted as a function of the mean $r^*$ magnitude. The small dots mark the
remaining 90569 sources with $-0.1 < g^* - r^* < 0.4$.
The two dashed lines show the boundary of the observational cutoff 
(the combination of items 1 and 2 above). The mean errors (for the magnitude 
difference) are about 0.03$^m$ up to 20$^m$, increasing to about 0.07$^m$ at 21$^m$, 
and to 0.2$^m$ at 22$^m$; the observed error distribution is in good 
agreement with the errors quoted by the photometric pipeline). 

There are 21 stars ($\about 10 \%$) rejected by the third condition
$|g^*_2-g^*_1| > |r^*_2-r^*_1|$, where indices 1 and 2 mark data from each epoch.
This condition is shown by the diagonal dashed lines in Figure 1b, where for clarity 
only stars with mean $r^* < 20$ are plotted (using the same symbols as in 
Figure 1a). Note that Figures 1a and 1b show two projections of the three-dimensional
selection volume spanned by $g^*_2-g^*_1$, $r^*_2-r^*_1$, and the mean $r^*$ 
magnitude. The rejected candidates are probably W UMa stars, but may also be RR Lyrae 
stars scattered across the selection boundary by photometric errors. The former 
probably dominate in the rejected subsample because more than $80\%$ are brighter 
than $r^*=20$, and thus have very small photometric errors.
The diagonal solid line in Figure 1b shows a best-fit relation  
$|g^*_2-g^*_1| = 1.4 \, |r^*_2-r^*_1|$. Similar analysis based on data from
other bands yields $|u^*_2-u^*_1| = 1.0 \, |g^*_2-g^*_1|$, 
$|r^*_2-r^*_1| = 1.2 \, |i^*_2-i^*_1|$, and $|i^*_2-r^*_1| = 1.1 \, |z^*_2-i^*_1|$, 
where the uncertainty of best-fit coefficients is $\la 0.05$.

A striking feature in Figure 1a is the lack of faint objects with large variability 
amplitudes in the regions outlined by the two 
ellipses. The magnitude distribution of the RR Lyrae candidates turns off
rather sharply at r$^* \sim$ 20, even though nonvariable objects are detected  
2.5 magnitudes fainter, and the errors are sufficiently small 
that variable sources with similar amplitudes could be detected
to at least r$^* \sim$ 21 (see below). This indicates that we are detecting the 
faint end of the RR Lyrae magnitude distribution, and hence the limit 
of their distance distribution. The absence of faint variable sources appears 
not to be due to our selection criteria, since sources in those two 
regions are already practically absent in the starting sample (small dots in Figure 1a). 
To repeat, the starting sample was selected from the full data set by simply 
extracting unresolved\footnote{Ignoring the requirement that candidates must 
be unresolved has practically no effect on the resulting sample. Thus possible 
star-galaxy misclassification cannot be invoked as an explanation for the observed
cutoff.} sources in the appropriate $g^* - r^*$ color range.

Additional support for the reality of this cutoff comes from the 
analysis of the overlap between runs 77 and 745. These runs were obtained 
9 months apart, which is a sufficiently long baseline to detect quasar (QSO) 
variability. Although low-redshift QSOs ($z <$ 2) have $g^* - r^*$ 
colors similar to RR Lyrae stars, they are easily distinguished 
by their bluer $u^*-g^*$ colors (\cite{Fan99}). Since variable QSOs 
should not have a faint magnitude cutoff, they can be used to 
test whether the data allow the detection of variable sources fainter 
than r$^* \sim$ 20. 

We searched for variable objects in the overlap of runs 77 and 745 
($\sim$ 35 deg$^2$) using the same criteria outlined above. This new search is 
summarized in the right column panels in Figure 2, and compared to the 
results from the first search shown in the panels on the left. The top 
two panels are analogous to Figure 1a (indeed, the top left panel is the same). 
Vertical lines at $r^* = 20.2$ are added to guide the eye, and mark the apparent faint 
end of the RR Lyrae magnitude distribution detected in runs 745-756. QSOs 
are selected by requiring $u^*-g^* < 0.8$ and are marked by 
crosses\footnote{Detailed analysis of the variable QSO sample is outside 
the scope of this work and will be presented in a separate publication.}. 
The color difference between QSOs and RR Lyrae stars can be easily seen in 
the $g^* - r^*$ vs. $u^* - g^*$ color-color diagrams shown in the two middle 
panels\footnote{In all color-color diagrams, blue is towards the lower left corner 
and red is towards the upper right corner.}. In these diagrams 
variable sources are marked by lines which connect photometric 
measurements at the two epochs, and dots represent a subsample of 
5000 nonvariable unresolved objects which outline the stellar locus and
the position of low-z QSOs ($u^*-g^* < 0.8$). Note that there are no 
variable sources with $u^*-g^* < 0.8$ in the left panel since QSOs do 
not vary much on a two day timescale. The lower two panels display $r^*$ 
vs. $g^* - r^*$ color-magnitude diagrams and show that we detect no RR 
Lyrae stars fainter than $r^* \sim 20$ even though variable QSOs, which 
are selected by identical criteria, and without using the $u'$ band data,
are detected to $r^* > 21$.

The described search procedure deliberately used only $g'$ and $r'$ band 
data in the first step in order to simplify selection effects, and thus
to show that the faint magnitude limit of selected \RRc\ is
real. For subsequent analysis, however, we further
constrain the sample to sources with colors appropriate for RR Lyrae stars 
($1.0 <u^*-g^* < 1.5$, $-0.1 <g^*-r^* < 0.4$,
$-0.2 <r^*-i^* < 0.2$, $-0.2 <i^*-z^* < 0.2$). These limits are shown as
boxes in the color-color diagrams displayed in Figure 3 and result in the
final sample of 148 stars. Thus, 80\% of the sources in the 
initial sample of 186 stars pass these tight additional criteria, showing
that \RRc\ can be quite efficiently selected with only two-epoch 
two-band  ($g'$ and $r'$) data with a sufficiently short baseline to avoid 
contamination by variable QSOs. About two-thirds of the 38 rejected sources 
narrowly fail one or two of the imposed color limits, and the remaining
third usually fails by more than 0.5 magnitudes in a single color,
most often in $u^*-g^*$ and $i^*-z^*$. While the former may be RR Lyrae 
stars, the latter are more likely to belong to various types of variable 
binary stars. The inclusion of the rejected sources in the subsequent analysis 
does not significantly change the resulting volume density of the selected
candidates. However, it affects the estimated statistical significance 
of the observed cut-off in the candidates' magnitude distribution at
$r^* \about 20$ (\S 4.1).

\section{      Analysis of the Candidate RR Lyrae Stars   }

The colors and variability properties of 148 stars in our final 
sample are consistent with their being RR Lyrae stars (\cite{KMS98}). 
In particular, our observations were obtained $\sim$ 2 days apart and 
thus are sensitive to the variability time scales characteristic for 
these stars (0.3-0.8 days, \cite{S84}), while insensitive to objects
varying on longer timescales (QSOs, long-period variables, etc.). 
In addition, the brightness variations are consistent with RR Lyrae 
amplitudes (0.7-1.5 mag peak-to-peak, \cite{S84}), and the candidates are 
bluer at the brighter epoch. While contamination by variable stars of similar 
properties (e.g. dwarf Cepheids or SX Phe stars) cannot be excluded without 
detailed light curves and/or spectroscopic data, the expected level 
of contamination is probably not larger than 10\% (\cite{H93}, \cite{G94}, 
see also \S4.3). We will assume in the rest of this work that all
sources in the final sample are RR Lyrae stars.

The selected \RRc\ are not smoothly distributed in magnitude, as can be seen 
in Figures 1 and 2. This clumpiness is also seen in their angular 
distribution. Figure 4 shows the distribution of \RRc, marked as open circles, 
in the mean $r^*$ vs. RA diagram. There is an obvious concentration of $\sim$ 
70 sources with $r^* \sim 19-19.5$ and 205$^\circ$ $<$ RA $<$ 230$^\circ$. This 
feature is present in all six data columns (separated in declination), and the 
column-to-column scatter of counts in these RA and $r^*$ ranges is consistent 
with Poisson statistics. 

To test this feature further, we analyze additional two-epoch data from several 
different run combinations. As before, the detection of variable QSOs in the
overlap of runs 77 and 745 allows a powerful test because variable QSOs should 
not display any spatial structure on such a large angular scale ($\ga 10^\circ$). 
The distribution of variable objects detected 
in these two runs in the mean $r^*$ vs. RA diagram is shown in Figure 5 (note 
the different RA limits here from those in Figure 4). Sources with $u^*-g^* > 
0.8$, \RRc, are marked by open circles and the sources with $u^*-g^* < 0.8$, presumably 
variable QSOs, are marked by solid squares. While QSOs are homogeneously 
distributed as expected, the distribution of \RRc\ is markedly 
different even though both samples were selected by identical criteria. 
At the same time the distribution of \RRc, most notably the 
concentration of sources with $r^* \sim 19-19.5$ at 215$^\circ$ $<$ RA $<$ 230$^\circ$, 
is in agreement with the results obtained from the overlap of runs 745 and 756, 
and shown in Figure 4. An overdensity of \RRc\ in the same magnitude-RA region  
is also detected in overlaps from runs 77-756 and 752-756.

A stellar overdensity analogous to that of \RRc\ in the region
205$^\circ$ $<$ RA  $<$ 230$^\circ$ can also be seen in the distribution of 
nonvariable sources with similar colors. Figure 6 displays the mean $r^*$ vs. 
RA diagram for 587 stars from the overlap of runs 745 and 756 satisfying 
$1.1 < u^*-g^* < 1.5$ and $-0.1 < g^*-r^* < 0.3$. The concentration of sources 
with $r^* \sim 19-19.5$ and 205$^\circ$ $<$ RA $<$ 230$^\circ$, the same magnitude--RA 
range as for the overdensity of \RRc\ displayed in Figure 4, is easily discernible 
and provides additional evidence for the clump. We used somewhat tighter color
criteria than in the search for \RRc, in order to decrease contamination by F stars 
from the blue tip of the stellar locus (c.f. Figure 2). However, this contamination
cannot be entirely removed by using only color cuts, and sources with $r^* > 20$ 
displayed in Figure 6 are probably main sequence or blue straggler stars (i.e. 
with smaller luminosities than RR Lyrae stars, and thus intrinsically fainter).
Yanny et al. (2000a) also detect this clump and another smaller clump in the southern 
Galactic hemisphere in SDSS commissioning data by analyzing the distribution of stars 
with similar colors ($0.8 < u^*-g^* < 1.5$ and $-0.3 < g^*-r^* < 0.0$).

A group of sources at RA $\sim$ 230$^\circ$ and $r^* \sim 17.5$ can be seen in both
Figures 5 and 6. These sources belong to the globular cluster Palomar 5 
(\cite{Rosino}, \cite{Abell}, \cite{Wilson}) and represent its blue horizontal branch stars. 
Figure 7 displays an $r^*$ vs. $g^*-r^*$ color-magnitude diagram for $\sim 2000$  
stars observed in run 756 inside a circle with 5 arcmin radius and centered on the 
position of the Palomar 5 core (J2000 RA = 15$^h$ 16$^m$ 5.3$^s$, Dec = $-0^\circ$ 6' 41"). 
Five stars selected here as \RRc\ (from the overlaps 745-756, 77-745 and 77-756) 
are marked by lines which connect measurements at different epochs, and all 
fall in the appropriate blue horizontal branch region for Palomar 5 ($g^* \sim$ 17.5, 
\cite{Smith86}). This further reinforces the assumption that our selection criteria 
reliably select RR Lyrae stars.

\subsection {    Galactic Distribution of \RRc    } 

The observed magnitudes of \RRc\ can be used to infer their distances and 
consequently their Galactic distribution. 
We calculate distances to stars in the final sample by assuming 
constant luminosity of M$_V$ = 0.7$^m$   
(\cite{L96}) and transformation M$_V$ = M$_{r^*} + 0.44(g^* - r^*) - 0.02$ 
(\cite{KMS98}) which typically results in  M$_V$--M$_{r^*}$ $\sim$ 0.05$^m$. 
For the apparent brightness estimate we use the mean $r^*$ 
magnitude corrected for the interstellar extinction, and for a 0.1$^m$ bias 
due to asymmetric RR Lyrae light curves (see Appendix A). Typical values of 
the interstellar extinction, as determined from the maps given by 
Schlegel, Finkbeiner \& Davis (1998), are $A_{r^*} = 0.05-0.15$ ($A_{g^*}$ = 
1.38 $A_{r^*}$).  

Figure 8 shows the Galactic distribution of \RRc, displayed as small circles
(the large circle marks the Sun's position at X=$-8$ kpc, Y=0, Z=0). The dashed 
lines show the volume within which our data can detect RR Lyrae stars: a very 
thin wedge with an opening angle of 80 deg and distances ranging from 5 kpc 
(saturation limit, $r^* \sim 14$) to 90 kpc (faint limit, $r^* \sim 21$).
The dotted lines show the intersection of this wedge 
with a galactocentric sphere of radius 30 kpc ($r^* \sim 18-18.5$).

The clump of \RRc\ centered on (X=20 kpc, Y=10 kpc, Z=40kpc), 
corresponds to the concentration of sources with $r^* \sim 19-19.5$ and 
205$^\circ$ $<$ RA $<$ 230$^\circ$ visible in Figure 4. The clump center is in a 
similar direction (l=340$^\circ$, b=60$^\circ$) as the center of the Sagittarius dwarf 
spheroidal galaxy at (l=5.6$^\circ$, b=$-14.0^\circ$). The Sagittarius dwarf spheroidal 
is the closest known Galactic satellite, with a galactocentric distance of 16$\pm$2 kpc
(\cite{I97}). In the coordinate system displayed in Figure 8 it is situated at 
X=15 kpc, Y=$-2$ kpc, Z=$-6$ kpc and marked by a triangle in the middle panel. 
The distance between its center and the clump is $\sim$ 50 kpc. This is 
significantly larger than the extent of either structure ($\sim$ 10 kpc) and 
probably implies that they are not physically associated. It is interesting, 
however, that the Galactic orbit of the Sgr dwarf spheroidal, as calculated by 
both Ibata {\em et al.} and Johnston {\em et al.} (1999a) crosses the clump
of \RRc. We display this orbit, taken from Johnston {\em et al.} (1999a), by 
a solid line in Figure 8 (the direction of Sgr dwarf's motion is towards the clump).
Such close proximity between the calculated orbit and the clump of \RRc\ may 
perhaps be evidence for presumed debris caused by tidal disruption of the Sgr 
dwarf spheroidal (\cite{J99a}, \cite{J99b}, Ibata {\em et al.} 2000) in the 
Galactic potential. Additional observations of the surrounding area, and the 
radial velocity measurements for the clump stars, are required to further explore this 
hypothesis.

From the Galactic distribution of \RRc\ presented in Figure 8, we calculate
their volume density as a function of galactocentric radius, and display it
as data points with 1 $\sigma$ error bars in Figure 9. The uncertainties are 
determined from Poisson statistics in the vertical direction and from the bin 
width in the horizontal direction. Since two-epoch data cannot detect all 
RR Lyrae stars, the overall normalization of the volume density includes unknown 
selection efficiency. We estimate that the selection efficiency is 56\%, from 
a Monte Carlo study based on a set of model light curves with realistic 
amplitude and period distributions (for details see Appendix A). This
estimate agrees well with two independent determinations described in \S4.3 
below.

Wetterer \& McGraw (1996) used a large compilation of available RR Lyrae 
searches to find that their distribution follows an $R^{-3}$ power law,
where $R$ is the Galactocentric radius. This is plotted as the thin solid line 
in Figure 9; the thin dot-dashed lines represent their 1 $\sigma$ normalization
uncertainty ($\sim$ factor of 2). We find two noteworthy deviations 
from this power law. First, our analysis indicates that the RR Lyrae volume density 
may follow a shallower power law with a best-fit index of 2.7 $\pm$ 0.2. Second, 
the absence of RR Lyrae stars with $r^* >$ 20 implies a rather sharp halo 
edge at $R_{halo}$=50-60 kpc. Figure 9 shows two power-law fits: the thick 
dot-dashed line is the $R^{-2.7}$ power law determined for data with $R <$ 60 kpc, 
and the thick dashed line is the steep $R^{-11.2}$ power law determined for data 
with $R$ $>$ 50 kpc. The latter power-law is shown only for illustration and 
should not be taken literally (note that there are only two data points at 
$R$ $>$ 50 kpc). Note that this sample is neither large enough, nor sufficiently  
extended over the sky, to constrain the halo flattening (e.g. Hartwick 1999).

These results imply that there are 2-3 times more RR Lyrae stars 
at $R \sim {\rm R}_{halo}$ than predicted by the Wetterer \& McGraw power law.
However, it is obvious that a shallower power-law is obtained mainly 
because of the large number of candidates in the ``45 kpc'' clump.
When they are excluded by constraining the sample to radii less than 
35 kpc, the best fit power-law fit becomes $R^{-3.1\pm0.2}$, in agreement
with the Wetterer \& McGraw result, including the normalization at the bright
end. The same result is obtained when the sample is constrained to 
160$^\circ$ $<$ RA $<$ 200$^\circ$ (see Figure 4), and in this case the fit is 
satisfactory all the way to R $\sim$ 50 kpc. This may indicate that the halo 
contains clumpy overdensities inhomogeneously distributed within an underlying 
smooth $R^{-3}$ density distribution.

\section{                       Discussion                     }

\subsection { Statistical Significance of the Observed Cut-off at R $\sim$ 50 kpc }

The statistical significance of the observed lack of candidate RR Lyrae stars
with $r^* > 20$ can be determined from the expected number of candidates
with such magnitudes. However, it is difficult to estimate this number with 
certainty because of the observed clumpy spatial distribution of selected 
candidates, and because of only weakly constrained selection effects at the 
faint end. We estimate the expected number of candidates with $r^* > 20$ by
extrapolating the $R^{-3}$ power-law density to infinity (this power-law 
implies a flat magnitude distribution for sources with uniform luminosity).

A Monte Carlo study of the selection effects described in the 
Appendix finds that the selection efficiency starts to fall off slowly for 
$r^* > 19$, reduces to about 50\% at $r^* \about 20$, and drops to 0 for 
$r^* \about 21.5$. Confirmation of this fall-off comes from 
analyzing the variable QSO sample. Assuming that the fraction of variable QSOs 
does not depend on their magnitude (we find this fraction to be $\about$ 15\%), 
the efficiency determined from simulations is in agreement with that implied
by the observed numbers of variable QSOs (which, however, may have very different 
light curves). We adopt 25\% as the mean efficiency in the $20 < r^* < 21.5$ 
magnitude range. In order to avoid the effects of the clumpy spatial distribution, \
we determine the expected density of candidates by considering only those 
satisfying $160^\circ < {\rm RA} < 200^\circ$ and $15 < r^* < 19$ (c.f. Figure 4). 
We estimate a density of 18 mag$^{-1}$, normalized to the entire RA  range. 
With the adopted efficiency, we find that the expected number of candidate
RR Lyrae stars with $r^* > 20$ is 7, while we have selected none. The Poisson 
probability for this outcome is $\about 10^{-3}$.

A more conservative approach may be taken by considering all 186 candidates
from the first selection step (i.e. before the color cuts described in \S 2 
were imposed to yield the final sample of 148 sources). Due to a larger number
of candidates, the expected number of sources with $r^* > 20$ is increased to 9. 
As can be seen in Figures 1a and 2, there are 4 sources which barely missed the 
cuts and could perhaps be RR Lyrae stars. The Poisson probability that 4 or fewer
sources are observed, given the expectation value of 9, is $\about 0.02$.
As evident, this approach significantly reduces the implied statistical significance 
of the observed cut-off at $r^* \about 20$. 

We conclude that the significance of the observed cut-off is at the level 
of 2-3 $\sigma$ (in terms of equivalent Gaussian probability). The best way to 
improve this estimate is to obtain follow-up observations of the rejected sources 
to establish whether they are RR Lyrae stars, and of course to analyze data for a 
significantly larger sky area.

\subsection {       Comparison with FASTT and LONEOS data      }

The final sample of \RRc\ presented here is based on commissioning data
taken at only two epochs. In order to estimate the level of spurious 
variability detections we have cross-referenced our list with 
the list of variable objects found from the FASTT (Flagstaff 
Astrometric Scanning Transit Telescope) data (\cite{HS98}) and 
with the LONEOS (Lowell Observatory Near Earth Object Search\footnote{LONEOS 
homepage is http://asteroid.lowell.edu/asteroid/loneos/loneos\_disc.html}) 
database. The LONEOS data fully cover the overlap of runs 745 and 756, 
while the overlap with the FASTT fields is only partial ($\sim$ 40 deg$^2$).
Both data sets can detect variable stars brighter than $r^* \sim 17$.

Henden \& Stone (1998) require that a candidate variable star show night-to-night
scatter (typically 8-12 epochs) larger than 3 times the expected error 
in the magnitude. Based on a random sampling of a set of model light 
curves with realistic amplitude and period distributions (see Appendix A),
we determine that the mean root-mean-square scatter for RR Lyrae stars 
is $\sim$ 0.3$^m$. The photometric accuracy of FASTT data (see Figure 1a in 
\cite{HS98}) implies that the detection efficiency for RR Lyrae stars 
in FASTT data should drop sharply for $r^* > 17^m$. The list of
selected \RRc\ includes 16 sources which are sufficiently bright 
and in the FASTT fields. We find that 14 of them are indeed in the Henden \& Stone
list as published, and one was detected later from additional observations 
(A. Henden, priv. comm.). This implies a reliability of our selection procedure 
of 94\%. The remaining source, with a mean $r^*=14.9$ (SDSSp J110838.26$-$000514.3, 
see Table 2 below), is definitely variable in SDSS data, with amplitudes exceeding 
0.15$^m$ in all 5 bands (0.34$^m$ in $g^*$). To estimate the efficiency of our 
two-epoch selection procedure, we matched all sources with appropriate colors 
(c.f. \S 2) detected in 
the SDSS data (8786 sources) to the Henden \& Stone list and found 33 matches 
(note that here no magnitude limit is imposed). As we recovered 17 of them as \RRc, 
this implies an efficiency of (52$\pm$15)\%, assuming that all Henden \& Stone variable 
sources with appropriate SDSS colors are RR Lyrae stars.

The sensitivity of the LONEOS data for variability detection drops sharply 
for $r^* > 17^m$, and 33 stars in our sample are brighter 
than this limit. For two of these stars no LONEOS data are available (both 
stars are too close to much brighter objects). Of the remaining 31 stars 
the LONEOS data clearly indicate variability for 29 (94\%) and
are inconclusive for the other two. Unfortunately, because of their 
coarse sampling in time, neither the LONEOS data nor the FASTT data can be 
used to produce light curves which could definitively identify the variable 
objects as RR Lyrae stars.

\subsection {Contamination by Other Variable Stars }

While the SDSS multi-color photometry indicates that all candidates 
discussed here have colors appropriate for RR Lyrae stars, there exist other 
types of variable star with similar colors. For example, W UMa stars can 
have F type spectra but they should be efficiently screened out by the 
selection requirement that candidates must be brighter in $r^*$ when they 
are bluer in $g^* - r^*$.

The most significant contaminants are probably $\delta$ Scu stars, 
another pulsating variable type with approximately the same color range 
as the RR Lyrae stars, periods of 1-3 hours, and magnitudes about 0.5$^m$ 
fainter than RR Lyrae stars (Hoffmeister {\em et al.} 1985). For example, 
the Tycho photometric survey finds that for $m_B < 11^m$, the numbers of RR 
Lyrae stars and $\delta$ Scu stars are comparable. 
However, the Population I $\delta$ Scu stars should start to run out at 
a distance of $\sim$ 2 kpc from the Galactic plane, or equivalently at 
$r^* < 13^m$ for $b > 30^\circ$.  On the other hand, the Population II 
$\delta$ Scu stars, also known as SX Phe stars, contribute only 10\% 
to the $\delta$ Scu population (G. Burks, priv. comm.). Thus, the likely 
contamination of our sample should not be more than 10\%, which would 
not qualitatively change our results. We note that this fraction may
be somewhat larger if the number of SX Phe stars decreases more slowly 
with Galactocentric radius than does the number of RR Lyrae stars.  
In the following section we discuss multi-epoch photometric observations 
for seven SDSS-selected candidate RR Lyrae stars, and show that 
all observed light curves have shapes and periods typical for 
RR Lyrae stars. This result is consistent with the above estimate
for the contamination fraction of $\la$ 10\%.

Another independent piece of evidence that our sample is dominated by 
the low-gravity RR Lyrae stars is that the $u^*-g^*$ colors for the 
selected candidates are redder than those of the parent population. 
It has been long known 
that Balmer jumps are bigger for Horizontal Branch stars than for Main 
Sequence stars (e.g. Oke, Giver \& Searle 1962, Pier 1983). The resulting 
$u^*-g^*$ colors for Horizontal Branch stars 
are redder than for Main Sequence Stars, given the same $g^*-r^*$ color
(Lenz et al. 1998). Figure 10 shows the $u^*-g^*$ color
distribution for stars selected by $0.0 < g^*-r^* < 0.1$ (c.f. Figure 2,
this narrow range of $g^*-r^*$ color selects stars with similar effective
temperature, $\about$ 7000-8000 K, Lenz {\em et al.} 1998) 
from the 90569 stars shown in Figure 1a (dashed line) and for stars selected
by the same criterion from the resulting sample of candidate RR Lyrae stars 
(solid line). The $u^*-g^*$ color of candidate RR Lyrae stars is 0.2 mag redder, 
on average, than that of nonvariable stars within the same narrow range of 
$g^*-r^*$ color. For Main Sequence stars  the expected $u^*-g^*$ range is 
0.85--1.0 (indicated as a horizontal line marked as log(g)=4.5 in the figure) 
and for Horizontal Branch stars the expected $u^*-g^*$ range is 1.1--1.3 
(horizontal line marked as log(g)=2.5), as found by Lenz {\em et al.} 1998. 
The intrinsic spread of the $u^*-g^*$ color for a given 
gravity is due to varying metallicity; the plotted values correspond 
to the range $-2.0 < [{\rm M}/{\rm H}] < 0$. The observed distribution of the 
$u^*-g^*$ colors of nonvariable stars is consistent with them being 
a mixture of low-gravity and high-gravity stars (for a detailed study of
A stars detected in SDSS commissioning data see Yanny {\em et al.} 2000a). 

The number of candidates found here agrees within 1$\sigma$ with the 
normalization given by Wetterer \& McGraw (1996). While this supports the
low level of contamination by other types of variable star,
the large uncertainty of their normalization and our efficiency prevents 
an accurate determination of such contamination in our sample. 
As already discussed in \S3, our 
normalization assumes an efficiency for detecting RR Lyrae stars from 
two-epoch data of 56\% determined by the procedure described in Appendix A. 
This estimate agrees well with the efficiency determined from the comparison 
with the FASTT data (52$\pm$15 \%). Another way to estimate efficiency
is to compare the subsamples detected in the 20.9 deg$^2$ overlap of 
runs 77, 745 and 756. There are 41 candidates selected from runs 745 and 
756, and 37 candidates from runs 77 and 745. These two subsamples have
16 sources in common, implying an efficiency of (41$\pm$12)\%, in 
good agreement with the above estimates.

\subsection { Light Curves for a Subsample of Candidate RR Lyrae Stars }

The accepted identifying characteristics of an RR Lyrae star are its light
curve and period. Here we present preliminary results of follow-up observations 
for several candidate RR Lyrae stars selected from SDSS commissioning data.

One of the faintest candidates in the sample, SDSSp J113049.26$-$005918.2 
($r^* \about 19.4$), was monitored for five nights during March 2000 with the 
SDSS 20-inch photometric telescope at Apache Point Observatory, New Mexico. 
25 individual measurements in $g'$ band were taken over the five nights; 
the resulting light curve is shown in Figure 11a. The preliminary 
estimate for the period is 0.46379 d, and the the first maximum is at 
HJD 2451606.2323. Both the light curve shape and the period confirm that this 
candidate is a bona fide RR Lyrae star. The observed apparent magnitude of 
this star places it at $\about$ 55 kpc from the Galactic center, and at $\about$ 
44 kpc from the Galactic plane. 

Another sample of six candidate RR Lyrae stars was selected from SDSS 
commissioning data obtained in the southern Galactic hemisphere 59 days apart
during the fall of 1998. While these data are not used in this work, the 
candidates were selected by the identical selection procedures. The candidates 
were observed in the Johnson V band by the 0.76-m reflector of the University of 
Washington's Manastash Ridge Observatory during the fall of 1999. These
observations were supplemented with data obtained from the LONEOS database
and the resulting light curves are shown in Figure 11b, together with the
candidates' names and periods. All six candidates have light curves with 
shapes and periods characteristic of RR Lyrae stars. Their average
Galactocentric distance is 30 kpc. 

These prefatory follow-up observations show that the majority of stars 
in our sample are probably RR Lyrae stars, and support the estimate
that the contamination by other types of variable star is $\la$ 10\%.

\subsection{               Future Work                  }

These preliminary results obtained with a small sample of SDSS data indicate
its potential for various Galactic structure studies. For example, 
the SDSS-FASTT and SDSS-LONEOS comparisons, and the light curves presented in
Figure 11, demonstrate that even two-epoch SDSS photometric data are sufficient 
for efficient detection of variable stars. The sample of \RRc\ presented here 
shows that such a deep and wide area survey may significantly contribute to 
studies of the outer Galactic halo. Nevertheless, it is probable that this sample 
does not contain only RR Lyrae stars, and that 
some of the assumptions in our analysis may not be valid. The most straightforward 
approach to determine the contamination level by variable stars other 
than RR Lyrae stars is to obtain light curves for all candidates (so far only 
a few hundred, but over the next few years SDSS will produce several thousand 
\RRc). To facilitate such observations, Table 2 lists coordinates and two-epoch 
SDSS photometry in $g'$ and $r'$ bands for 148 candidates discussed in 
this work\footnote{The finding charts are available from the authors upon request.}. 
For the confirmed RR Lyrae stars the measurements of their radial velocity
offer the exciting possibility of measuring the distribution of dark matter
throughout the halo (\cite{H84}).

\vskip 0.4in
\leftline{Acknowledgments}

We are grateful to an anonymous referee for many insightful comments.
We also thank Geoff Burks, Bohdan Paczynski, and Christophe Alard for helpful 
discussions regarding the contamination of RR Lyrae sample by other variable 
stars.

The Sloan Digital Sky Survey (SDSS) is a joint project of The University of
Chicago, Fermilab, the Institute for Advanced Study, the Japan Participation 
Group, The Johns Hopkins University, the Max-Planck-Institute for Astronomy, 
Princeton University, the United States Naval Observatory, and the University of 
Washington. Apache Point Observatory, site of the SDSS, is operated by the 
Astrophysical Research Consortium. Funding for the project has been provided by 
the Alfred P. Sloan Foundation, the SDSS member institutions, the National 
Aeronautics and Space Administration, the National Science Foundation, the U.S. 
Department of Energy and Monbusho, Japan. 
The SDSS Web site is http://www.sdss.org/.

\appendix {\bf Appendix A: Monte Carlo Study of the Selection Effects }

To determine the sensitivity of two-epoch selection of RR Lyrae stars
presented in this work, we randomly sample a set of model light curves
with realistic amplitude and period distributions. We use for this purpose
a set of 180 RR Lyrae star light curves extensively measured by the ROTSE project
\footnote{see http://www.umich.edu/$\sim$rotse} as described by
Akerlof {\em et al.} (2000). Spline fits to these phased light curves provide our 
templates.

For each template object we sample the light curve at two points separated 
by the nominal 1.99462 days spacing for 1000 random phases. Because
this spacing is at least 3 times longer than typical RR Lyrae star periods 
($\la$ 0.7 days), the details of the adopted period distribution have only a
minor impact on the model results. For the same reason, stars with periods
shorter than RR Lyrae periods (e.g. SX Phe stars) are selected with roughly
the same efficiency. 
For each sampling we obtain a real magnitude difference, apply SDSS photometric 
errors to the two `measurements', and determine whether this object would pass our
variability selection criteria. This allows us to realistically assess our
efficiency as a function of magnitude. In addition it allows us to
characterize the effect of two epoch observations on our observed mean
magnitudes, and hence on distance estimates.

We find that at bright magnitudes, the detection efficiency is constant at 
the level of 56\%, which is imposed by the combination of the two-epoch
selection and RR Lyrae light curve shapes. The efficiency falls off slowly 
beginning at $r^* \sim 19$ due to increased photometric errors, and is reduced 
to half its peak value at $r^* \sim 20$. We also find that the mean magnitudes 
calculated from two-epoch data are biased towards the bright side for about 0.1$^m$ 
because the RR Lyrae light curves are not symmetric around the mean brightness. 
The mean root-mean-square deviation of RR Lyrae light curves used in this
analysis is 0.21$^m$, and is practically independent of the number of
epochs as long as it exceeds $\sim 10$. We note that the light curves
used in this analysis were obtained with an open CCD and thus are representative
of the red bands. Because of this effect, the RR Lyrae amplitudes in
bluer bands (e.g. $g'$ and $r'$) may be somewhat larger and we adopt a
conservative upper limit of 0.3$^m$ used in the comparison with the 
FASTT data (\S 4.1).


\newpage
\begin{scriptsize}
\begin{deluxetable}{rcccc}
\tablenum{1}
\tablecolumns{5}
\tablewidth{240pt}
\tablecaption{Overlaps Between SDSS Commissioning Runs.}
\tablehead
{
    Runs  & min RA$^a$    & max RA$^b$    & Area$^c$ &  $\Delta t^d$  
}
\startdata
 745-756  & 10$^h$ 42$^m$ & 15$^h$ 46$^m$ &   97.5   &  1.9946 \\   
  77-745  & 14$^h$ 10$^m$ & 16$^h$ 39$^m$ &   33.7   &  266.10 \\
  77-756  & 14$^h$ 10$^m$ & 15$^h$ 46$^m$ &   20.5   &  268.09 \\
 752-756  & 09$^h$ 41$^m$ & 15$^h$ 46$^m$ &   11.1   &  0.9976 \\ 
\enddata
\tablenotetext{a}{The starting RA for the overlap.}
\tablenotetext{b}{The ending RA for the overlap.}
\tablenotetext{c}{Total area in the overlap (deg$^2$).}
\tablenotetext{d}{Time elapsed between the two observations (in days).}
\end{deluxetable}
\end{scriptsize}

\newpage
\begin{scriptsize}
\begin{deluxetable}{rcccccccrr}
\tablenum{2}
\tablecolumns{10}
\tablewidth{470pt}
\tablecaption{Candidate RR Lyrae Stars from SDSS Runs 745-756.}
\tablehead
{
No. & SDSS Name & HS Name$^a$ & $g_1^*$ & $r_1^*$ & $g_2^*$ & $r_2^*$ & 
$\langle r^{*}\rangle^b$ & $r_2^*$ - $r_1^*$ & $\langle g^*-r^*\rangle^c$
}
\startdata
1  & SDSSp J104902.61+010500.6 & \othername & 17.66 & 17.38 & 16.75 & 16.78 & 17.08 &  $-$0.60 &  0.03  \\
2  & SDSSp J105314.69+011201.4 & H01010306 & 14.77 & 14.75 & 14.53 & 14.55 & 14.65 &  $-$0.20 & $-$0.01  \\
3  & SDSSp J105926.11$-$005927.6 & \othername & 17.49 & 17.44 & 18.26 & 18.01 & 17.73 &   0.57 &  0.08  \\
4  & SDSSp J110035.99$-$003315.9 & \othername & 17.98 & 17.96 & 18.23 & 18.20 & 18.08 &   0.24 & $-$0.04  \\
5  & SDSSp J110838.26$-$000514.3 & \othername & 14.98 & 14.80 & 15.32 & 15.01 & 14.91 &   0.21 &  0.07  \\ \\
6  & SDSSp J111010.79+010732.9 & \othername & 17.26 & 17.21 & 17.49 & 17.39 & 17.30 &   0.18 & $-$0.04  \\
7  & SDSSp J111705.98$-$003424.0 & \othername & 17.18 & 17.16 & 17.87 & 17.60 & 17.38 &   0.44 &  0.09  \\
8  & SDSSp J112425.37$-$000919.7 & \othername & 17.46 & 17.41 & 17.67 & 17.61 & 17.51 &   0.20 &  0.00  \\
9  & SDSSp J112837.73$-$000112.6 & \othername & 19.24 & 18.96 & 18.40 & 18.38 & 18.67 &  $-$0.58 &  0.02  \\
10  & SDSSp J113049.26$-$005918.2 & \othername & 19.84 & 19.63 & 18.92 & 18.92 & 19.27 &  $-$0.71 &  0.06  \\ \\
11  & SDSSp J113814.16+010528.2 & \othername & 19.23 & 18.98 & 18.78 & 18.68 & 18.83 &  $-$0.30 &  0.04  \\
12  & SDSSp J114542.24+002314.6 & \othername & 17.59 & 17.43 & 17.96 & 17.67 & 17.55 &   0.24 &  0.10  \\
13  & SDSSp J114602.26+002057.7 & \othername & 20.25 & 20.00 & 19.52 & 19.54 & 19.77 &  $-$0.46 &  0.04  \\
14  & SDSSp J115113.99+004505.7 & \othername & 15.97 & 15.72 & 16.17 & 15.92 & 15.82 &   0.20 &  0.05  \\
15  & SDSSp J115534.40$-$003601.9 & \othername & 16.83 & 16.66 & 17.03 & 16.82 & 16.74 &   0.16 &  0.09  \\ \\
16  & SDSSp J115628.60+011223.9 & \othername & 17.17 & 17.23 & 17.69 & 17.61 & 17.42 &   0.38 &  0.00  \\
17  & SDSSp J115706.95$-$005507.9 & \othername & 17.29 & 17.37 & 18.30 & 18.03 & 17.70 &   0.66 &  0.04  \\
18  & SDSSp J115724.21$-$005358.2 & \othername & 18.34 & 18.26 & 18.79 & 18.53 & 18.40 &   0.27 &  0.06  \\
19  & SDSSp J120047.92+004611.1 & \othername & 17.57 & 17.48 & 17.28 & 17.26 & 17.37 &  $-$0.22 &  0.05  \\
20  & SDSSp J120730.94$-$000412.6 & \othername & 16.97 & 16.95 & 17.81 & 17.60 & 17.27 &   0.65 &  0.03  \\ \\
21  & SDSSp J121329.64$-$010151.9 & \othername & 17.46 & 17.32 & 17.74 & 17.51 & 17.41 &   0.19 &  0.09  \\
22  & SDSSp J121507.76+004930.1 & \othername & 17.70 & 17.72 & 18.16 & 18.08 & 17.90 &   0.36 &  0.04  \\
23  & SDSSp J121527.79$-$005256.5 & \othername & 15.50 & 15.54 & 16.59 & 16.40 & 15.97 &   0.86 &  0.02  \\
24  & SDSSp J121803.72+001448.9 & \othername & 18.60 & 18.34 & 17.86 & 17.86 & 18.10 &  $-$0.48 &  0.04  \\
25  & SDSSp J122228.39$-$010216.3 & \othername & 15.23 & 14.99 & 14.98 & 14.77 & 14.88 &  $-$0.22 &  0.09  \\ \\
26  & SDSSp J122501.92+011407.9 &  I01010312  & 16.12 & 15.98 & 16.49 & 16.28 & 16.13 &   0.30 &  0.05  \\
27  & SDSSp J122529.03+011420.8 & I01010394  & 15.39 & 15.40 & 15.91 & 15.70 & 15.55 &   0.30 &  0.05  \\
28  & SDSSp J123829.78+002001.6 & \othername & 18.51 & 18.54 & 18.98 & 18.90 & 18.72 &   0.36 &  0.03  \\
29  & SDSSp J124032.87$-$000312.9 & \othername & 17.51 & 17.19 & 17.25 & 16.99 & 17.09 &  $-$0.20 &  0.14  \\
30  & SDSSp J124046.56+005006.2 & \othername & 18.20 & 18.09 & 17.78 & 17.84 & 17.96 &  $-$0.25 &  0.01  \\ \\
31  & SDSSp J124136.64+011306.5 & \othername & 17.15 & 17.09 & 17.56 & 17.31 & 17.20 &   0.22 &  0.09  \\
32  & SDSSp J124224.91$-$001203.1 & I20060386  & 14.41 & 14.43 & 14.69 & 14.64 & 14.54 &   0.21 &  0.00  \\
33  & SDSSp J125028.28$-$000021.8 & \othername & 19.35 & 19.14 & 19.55 & 19.31 & 19.23 &   0.17 &  0.09  \\
34  & SDSSp J125208.73$-$002931.8 &  I28090270 & 15.61 & 15.44 & 16.05 & 15.83 & 15.63 &   0.39 &  0.10  \\
35  & SDSSp J125917.31+010240.4 & \othername & 19.57 & 19.25 & 18.81 & 18.78 & 19.02 &  $-$0.47 &  0.08  \\ \\
36  & SDSSp J131045.97$-$002621.9 & \othername & 19.40 & 19.26 & 19.11 & 19.10 & 19.18 &  $-$0.16 &  0.02  \\
37  & SDSSp J131117.74$-$003429.9 & \othername & 15.05 & 15.11 & 15.96 & 15.75 & 15.43 &   0.64 &  0.02  \\
38  & SDSSp J131717.78$-$003558.0 & \othername & 19.08 & 18.92 & 18.84 & 18.71 & 18.82 &  $-$0.21 &  0.07  \\
39  & SDSSp J131757.46$-$000818.8 & J21020002  & 17.40 & 17.15 & 17.02 & 16.91 & 17.03 &  $-$0.24 &  0.05  \\
40  & SDSSp J131806.63$-$003300.2 & J29020030   & 15.88 & 15.87 & 16.10 & 16.06 & 15.96 &   0.19 &  0.02  \\ \\
41  & SDSSp J132036.99+010945.4 & \othername & 18.68 & 18.72 & 19.40 & 19.15 & 18.93 &   0.43 &  0.07  \\
42  & SDSSp J132158.06+010659.3 & \othername & 18.96 & 18.99 & 19.16 & 19.14 & 19.06 &   0.15 &  0.00  \\
43  & SDSSp J132624.94$-$002612.1 & J21040274  & 16.75 & 16.69 & 17.23 & 17.04 & 16.87 &   0.35 &  0.08  \\
44  & SDSSp J132635.08+002034.8 & \othername & 17.69 & 17.75 & 18.51 & 18.41 & 18.08 &   0.66 &  0.03  \\
45  & SDSSp J132700.05$-$005456.7 & J37040583 & 15.29 & 15.04 & 14.84 & 14.80 & 14.92 &  $-$0.24 &  0.05  \\ \\
\enddata
\end{deluxetable}
\end{scriptsize}

\begin{scriptsize}
\begin{deluxetable}{rcccccccrr}
\tablenum{2}
\tablecolumns{10}
\tablewidth{470pt}
\tablecaption{Candidate RR Lyrae Stars from SDSS Runs 745-756 -- continued.}
\tablehead
{
No. & SDSS Name & HS Name$^a$ & $g_1^*$ & $r_1^*$ & $g_2^*$ & $r_2^*$ & 
$\langle r^{*}\rangle^b$ & $r_2^*$ - $r_1^*$ & $\langle g^*-r^*\rangle^c$
}
\startdata
46  & SDSSp J132745.54+001925.5 & \othername & 19.21 & 18.92 & 18.82 & 18.69 & 18.80 &  $-$0.23 &  0.09  \\
47  & SDSSp J133252.91+004622.6 & J01060321  & 15.82 & 15.84 & 16.39 & 16.15 & 15.99 &   0.31 &  0.04  \\
48  & SDSSp J133323.47$-$001159.5 & \othername & 17.98 & 17.74 & 17.41 & 17.30 & 17.52 &  $-$0.44 &  0.05  \\
49  & SDSSp J133552.24$-$003706.8 &  J29070157  & 15.86 & 15.55 & 15.61 & 15.36 & 15.46 &  $-$0.19 &  0.10  \\
50  & SDSSp J133748.90$-$005646.6 & \othername & 19.45 & 19.19 & 18.82 & 18.81 & 19.00 &  $-$0.38 &  0.00  \\ \\
51  & SDSSp J134142.51+004210.5 & \othername & 19.84 & 19.79 & 20.05 & 19.98 & 19.88 &   0.19 & $-$0.05  \\
52  & SDSSp J134452.37+003810.2 & \othername & 19.32 & 19.13 & 19.06 & 18.92 & 19.02 &  $-$0.21 &  0.07  \\
53  & SDSSp J134513.92+002240.0 & \othername & 17.00 & 17.08 & 17.31 & 17.32 & 17.20 &   0.24 & $-$0.03  \\
54  & SDSSp J134521.33$-$000147.4 & \othername & 14.12 & 14.01 & 14.80 & 14.51 & 14.26 &   0.50 &  0.03  \\
55  & SDSSp J134529.59+002156.3 & \othername & 19.52 & 19.25 & 18.86 & 18.79 & 19.02 &  $-$0.46 &  0.07  \\ \\
56  & SDSSp J134650.04+001659.7 & \othername & 18.96 & 18.91 & 18.45 & 18.54 & 18.73 &  $-$0.37 & $-$0.08  \\
57  & SDSSp J134854.94+004622.4 & \othername & 19.29 & 19.20 & 18.81 & 18.84 & 19.02 &  $-$0.36 & $-$0.05  \\
58  & SDSSp J135007.45$-$005638.3 & \othername & 18.95 & 18.74 & 18.58 & 18.49 & 18.61 &  $-$0.25 &  0.10  \\
59  & SDSSp J135009.13$-$003414.3 & \othername & 15.60 & 15.66 & 16.61 & 16.38 & 16.02 &   0.72 &  0.04  \\
60  & SDSSp J135156.26$-$005314.2 & \othername & 18.96 & 18.85 & 19.20 & 19.00 & 18.93 &   0.15 &  0.04  \\ \\
61  & SDSSp J135231.76+004350.9 & \othername & 17.48 & 17.45 & 18.22 & 17.99 & 17.72 &   0.54 &  0.04  \\
62  & SDSSp J135233.36$-$003336.9 & \othername & 19.36 & 19.17 & 19.57 & 19.36 & 19.27 &   0.19 &  0.12  \\
63  & SDSSp J135724.66$-$001028.9 & \othername & 19.07 & 19.05 & 19.27 & 19.23 & 19.14 &   0.18 &  0.00  \\
64  & SDSSp J135738.22+002055.6 & \othername & 19.08 & 18.83 & 18.78 & 18.63 & 18.73 &  $-$0.20 &  0.09  \\
65  & SDSSp J135824.04$-$002818.8 & \othername & 18.87 & 18.79 & 19.22 & 19.00 & 18.89 &   0.21 &  0.10  \\ \\
66  & SDSSp J135828.73+001248.5 & \othername & 19.09 & 18.81 & 18.60 & 18.49 & 18.65 &  $-$0.32 &  0.07  \\
67  & SDSSp J135955.38$-$002627.3 & \othername & 19.64 & 19.35 & 19.07 & 18.95 & 19.15 &  $-$0.40 &  0.07  \\
68  & SDSSp J140202.34+011243.5 & \othername & 20.02 & 19.90 & 19.62 & 19.59 & 19.74 &  $-$0.31 &  0.01  \\
69  & SDSSp J140206.54$-$002708.2 & \othername & 20.26 & 19.96 & 20.00 & 19.72 & 19.84 &  $-$0.24 &  0.07  \\
70  & SDSSp J140312.54+004802.7 & \othername & 19.33 & 19.08 & 19.57 & 19.28 & 19.18 &   0.20 &  0.12  \\ \\
71  & SDSSp J140606.77$-$003356.8 & \othername & 15.58 & 15.35 & 15.74 & 15.50 & 15.43 &   0.15 &  0.07  \\
72  & SDSSp J140641.30+010818.1 & \othername & 19.60 & 19.34 & 19.11 & 18.97 & 19.16 &  $-$0.37 &  0.07  \\
73  & SDSSp J140829.95$-$003751.1 & \othername & 19.04 & 18.80 & 18.66 & 18.55 & 18.68 &  $-$0.25 &  0.10  \\
74  & SDSSp J140849.79$-$000422.2 & \othername & 19.15 & 18.97 & 19.41 & 19.18 & 19.07 &   0.21 &  0.08  \\
75  & SDSSp J141059.99$-$002916.3 & \othername & 19.12 & 19.11 & 19.60 & 19.40 & 19.25 &   0.29 &  0.07  \\ \\
76  & SDSSp J141142.14+002248.5 & \othername & 16.41 & 16.13 & 15.86 & 15.73 & 15.93 &  $-$0.40 &  0.09  \\
77  & SDSSp J141238.55$-$005350.7 & \othername & 15.25 & 14.92 & 15.56 & 15.22 & 15.07 &   0.30 &  0.14  \\
78  & SDSSp J141446.16$-$002836.8 & \othername & 18.81 & 18.85 & 19.01 & 19.00 & 18.93 &   0.15 &  0.01  \\
79  & SDSSp J141543.43$-$000613.0 & \othername & 16.85 & 16.66 & 17.22 & 16.97 & 16.81 &   0.31 &  0.08  \\
80  & SDSSp J141554.95+011003.4 & \othername & 19.67 & 19.44 & 19.36 & 19.23 & 19.34 &  $-$0.21 &  0.05  \\ \\
81  & SDSSp J141724.61$-$000056.5 & \othername & 19.63 & 19.37 & 18.68 & 18.74 & 19.05 &  $-$0.63 &  0.05  \\
82  & SDSSp J141807.36+002302.6 & \othername & 15.33 & 15.10 & 14.92 & 14.81 & 14.96 &  $-$0.29 &  0.05  \\
83  & SDSSp J141846.15+010826.6 & \othername & 19.12 & 19.02 & 19.68 & 19.39 & 19.20 &   0.37 &  0.06  \\
84  & SDSSp J141858.06$-$002643.0 & \othername & 18.70 & 18.47 & 18.05 & 18.00 & 18.23 &  $-$0.47 &  0.06  \\
85  & SDSSp J141927.26+002215.7 & \othername & 19.83 & 19.53 & 19.60 & 19.37 & 19.45 &  $-$0.16 &  0.07  \\ \\
86  & SDSSp J141934.16$-$005509.4 & \othername & 19.78 & 19.48 & 19.55 & 19.26 & 19.37 &  $-$0.22 &  0.11  \\
87  & SDSSp J142112.29+003936.3 & \othername & 19.52 & 19.12 & 18.86 & 18.79 & 18.95 &  $-$0.33 &  0.02  \\
88  & SDSSp J142257.47$-$002922.8 & \othername & 19.50 & 19.26 & 19.03 & 18.98 & 19.12 &  $-$0.28 &  0.11  \\
89  & SDSSp J142321.66$-$000705.1 & \othername & 19.33 & 19.29 & 20.01 & 19.78 & 19.54 &   0.49 &  0.05  \\
90  & SDSSp J142337.07+002502.7 & \othername & 19.54 & 19.42 & 19.85 & 19.67 & 19.55 &   0.25 &  0.11  \\ \\
\enddata
\end{deluxetable}
\end{scriptsize}

\begin{scriptsize}
\begin{deluxetable}{rcccccccrr}
\tablenum{2}
\tablecolumns{10}
\tablewidth{470pt}
\tablecaption{Candidate RR Lyrae Stars from SDSS Runs 745-756 -- continued.}
\tablehead
{
No. & SDSS Name & HS Name$^a$ & $g_1^*$ & $r_1^*$ & $g_2^*$ & $r_2^*$ & 
$\langle r^{*}\rangle^b$ & $r_2^*$ - $r_1^*$ & $\langle g^*-r^*\rangle^c$
}
\startdata
91  & SDSSp J142356.74$-$003428.5 & \othername & 16.35 & 16.19 & 15.99 & 15.95 & 16.07 &  $-$0.24 &  0.07  \\
92  & SDSSp J142502.47$-$005331.7 & \othername & 19.56 & 19.41 & 19.27 & 19.24 & 19.32 &  $-$0.17 &  0.09  \\
93  & SDSSp J142530.26$-$005153.8 & \othername & 19.59 & 19.28 & 18.80 & 18.74 & 19.01 &  $-$0.54 &  0.04  \\
94  & SDSSp J142602.36+010837.7 & \othername & 20.02 & 19.74 & 19.40 & 19.37 & 19.55 &  $-$0.37 &  0.06  \\
95  & SDSSp J142742.51$-$002848.8 & \othername & 19.56 & 19.27 & 18.43 & 18.50 & 18.88 &  $-$0.77 &  0.04  \\ \\
96  & SDSSp J142807.14$-$000341.7 & \othername & 19.59 & 19.34 & 19.14 & 18.99 & 19.16 &  $-$0.35 &  0.11  \\
97  & SDSSp J142808.95$-$001148.3 & \othername & 16.51 & 16.34 & 16.09 & 16.12 & 16.23 &  $-$0.22 &  0.04  \\
98  & SDSSp J143032.31$-$000329.1 & \othername & 18.81 & 18.54 & 19.04 & 18.77 & 18.66 &   0.23 &  0.09  \\
99  & SDSSp J143241.74+001550.6 & \othername & 19.75 & 19.43 & 19.09 & 19.00 & 19.21 &  $-$0.43 &  0.05  \\
100  & SDSSp J143311.76+011356.3 & \othername & 19.45 & 19.25 & 19.68 & 19.41 & 19.33 &   0.16 &  0.12  \\ \\
101  & SDSSp J143312.89$-$000733.5 & \othername & 19.52 & 19.30 & 19.07 & 19.04 & 19.17 &  $-$0.26 &  0.04  \\
102  & SDSSp J143427.44$-$003721.2 & \othername & 19.01 & 19.02 & 19.35 & 19.28 & 19.15 &   0.26 &  0.03  \\
103  & SDSSp J143614.78+010825.9 & \othername & 14.81 & 14.81 & 15.17 & 15.08 & 14.95 &   0.27 &  0.04  \\
104  & SDSSp J143713.36+001623.0 & \othername & 15.72 & 15.71 & 16.09 & 15.87 & 15.79 &   0.16 &  0.05  \\
105  & SDSSp J143924.18$-$003211.9 & \othername & 18.77 & 18.77 & 19.09 & 18.94 & 18.86 &   0.17 &  0.07  \\ \\
106  & SDSSp J144003.43+001346.4 & \othername & 19.64 & 19.53 & 20.02 & 19.78 & 19.66 &   0.25 &  0.07  \\
107  & SDSSp J144424.47+010901.5 & \othername & 19.86 & 19.63 & 19.52 & 19.44 & 19.54 &  $-$0.19 &  0.04  \\
108  & SDSSp J144427.84$-$005806.4 & \othername & 19.33 & 18.99 & 18.79 & 18.79 & 18.89 &  $-$0.20 &  0.05  \\
109  & SDSSp J144618.52+001321.2 & \othername & 15.19 & 15.13 & 15.90 & 15.63 & 15.38 &   0.50 &  0.11  \\
110  & SDSSp J144720.41$-$000101.7 & \othername & 17.94 & 17.63 & 17.55 & 17.35 & 17.49 &  $-$0.28 &  0.10  \\ \\
111  & SDSSp J144939.60$-$002943.9 & \othername & 16.80 & 16.79 & 17.61 & 17.35 & 17.07 &   0.56 &  0.08  \\
112  & SDSSp J145258.17$-$000815.3 & \othername & 19.07 & 18.88 & 19.50 & 19.21 & 19.05 &   0.33 &  0.11  \\
113  & SDSSp J145414.56+002310.3 & \othername & 19.83 & 19.52 & 18.91 & 18.90 & 19.21 &  $-$0.62 &  0.06  \\
114  & SDSSp J145637.83$-$005622.8 & \othername & 18.88 & 18.92 & 19.88 & 19.61 & 19.27 &   0.69 &  0.06  \\
115  & SDSSp J145719.71$-$005328.0 & \othername & 16.06 & 15.76 & 15.52 & 15.37 & 15.56 &  $-$0.39 &  0.10  \\ \\
116  & SDSSp J150129.25$-$005433.3 & \othername & 19.21 & 18.84 & 18.75 & 18.62 & 18.73 &  $-$0.22 &  0.11  \\
117  & SDSSp J150147.85+004811.5 & \othername & 19.71 & 19.49 & 19.48 & 19.28 & 19.38 &  $-$0.21 &  0.09  \\
118  & SDSSp J150218.16$-$000947.9 & \othername & 19.51 & 19.38 & 19.16 & 19.17 & 19.27 &  $-$0.21 & $-$0.01  \\
119  & SDSSp J150257.73+001535.6 & \othername & 19.15 & 19.11 & 19.90 & 19.66 & 19.38 &   0.55 &  0.04  \\
120  & SDSSp J150337.35$-$002812.8 & \othername & 14.98 & 14.77 & 15.24 & 14.98 & 14.88 &   0.21 &  0.11  \\ \\
121  & SDSSp J150545.38$-$000505.3 & \othername & 16.20 & 16.14 & 16.97 & 16.66 & 16.40 &   0.52 &  0.09  \\
122  & SDSSp J150633.98+001806.6 & \othername & 17.49 & 17.42 & 17.79 & 17.71 & 17.57 &   0.29 & $-$0.01  \\
123  & SDSSp J150807.79$-$000300.3 & \othername & 19.62 & 19.52 & 20.05 & 19.92 & 19.72 &   0.40 &  0.04  \\
124  & SDSSp J150916.76+001947.2 & K14010591 & 16.09 & 16.06 & 16.76 & 16.49 & 16.27 &   0.43 &  0.06  \\
125  & SDSSp J151108.75$-$010015.2 & \othername & 19.72 & 19.46 & 18.81 & 18.89 & 19.18 &  $-$0.57 &  0.08  \\ \\
126  & SDSSp J151127.50$-$005511.9 & \othername & 19.56 & 19.54 & 19.98 & 19.87 & 19.70 &   0.33 &  0.00  \\
127  & SDSSp J151216.31$-$003643.0 & \othername & 19.44 & 19.24 & 19.67 & 19.39 & 19.31 &   0.15 &  0.16  \\
128  & SDSSp J151435.44$-$002959.7 & K27030069  & 15.27 & 15.27 & 15.64 & 15.50 & 15.38 &   0.23 &  0.01  \\
129  & SDSSp J151516.34$-$005124.2 & \othername & 19.96 & 19.65 & 19.67 & 19.41 & 19.53 &  $-$0.24 &  0.13  \\
130  & SDSSp J151557.21$-$000653.2 & \othername & 17.47 & 17.44 & 17.73 & 17.61 & 17.52 &   0.17 &  0.04  \\ \\
131  & SDSSp J151610.53+011410.9 & \othername & 20.08 & 19.80 & 19.57 & 19.38 & 19.59 &  $-$0.42 &  0.14  \\
132  & SDSSp J151659.66$-$005254.1 & \othername & 19.66 & 19.43 & 19.17 & 19.13 & 19.28 &  $-$0.30 &  0.07  \\
133  & SDSSp J151823.60+002122.2 & \othername & 17.70 & 17.42 & 17.35 & 17.27 & 17.34 &  $-$0.15 &  0.09  \\
134  & SDSSp J152014.18$-$002603.0 & K21041041  & 15.74 & 15.42 & 15.28 & 15.18 & 15.30 &  $-$0.24 &  0.12  \\
135  & SDSSp J152122.93$-$000530.9 & K21050106  & 17.06 & 17.10 & 17.90 & 17.60 & 17.35 &   0.50 &  0.08  \\ \\
\enddata
\end{deluxetable}
\end{scriptsize}

\begin{scriptsize}
\begin{deluxetable}{rcccccccrr}
\tablenum{2}
\tablecolumns{10}
\tablewidth{470pt}
\tablecaption{Candidate RR Lyrae Stars from SDSS Runs 745-756 -- continued.}
\tablehead
{
No. & SDSS Name & HS Name$^a$ & $g_1^*$ & $r_1^*$ & $g_2^*$ & $r_2^*$ & 
$\langle r^{*}\rangle^b$ & $r_2^*$ - $r_1^*$ & $\langle g^*-r^*\rangle^c$
}
\startdata
136  & SDSSp J152132.05$-$010202.9 & \othername & 19.56 & 19.45 & 19.34 & 19.27 & 19.36 &  $-$0.18 &  0.08  \\
137  & SDSSp J152318.61$-$005520.9 & K34050625  & 16.87 & 16.50 & 17.09 & 16.72 & 16.61 &   0.22 &  0.17  \\
138  & SDSSp J152547.10+002409.5 & \othername & 18.38 & 18.08 & 17.89 & 17.68 & 17.88 &  $-$0.40 &  0.11  \\
139  & SDSSp J152711.10+002506.4 & \othername & 19.44 & 19.35 & 19.88 & 19.68 & 19.52 &   0.33 &  0.02  \\
140  & SDSSp J152833.21$-$002546.9 & \othername & 19.83 & 19.66 & 19.58 & 19.48 & 19.57 &  $-$0.18 &  0.11  \\ \\
141  & SDSSp J153006.75+010806.7 & \othername & 20.02 & 19.75 & 19.49 & 19.33 & 19.54 &  $-$0.42 &  0.11  \\
142  & SDSSp J153129.00+001724.5 & \othername & 19.89 & 19.53 & 19.04 & 19.05 & 19.29 &  $-$0.48 &  0.04  \\
143  & SDSSp J153439.14$-$002615.4 & \othername & 19.24 & 19.09 & 19.52 & 19.30 & 19.20 &   0.21 &  0.11  \\
144  & SDSSp J153443.29$-$002937.9 & K27090033  & 15.57 & 15.46 & 16.15 & 15.82 & 15.64 &   0.36 &  0.13  \\
145  & SDSSp J153502.96+001421.5 &   K14090198 & 16.00 & 15.86 & 15.61 & 15.61 & 15.73 &  $-$0.25 &  0.01  \\ \\
146  & SDSSp J153518.04+001405.9 & \othername & 17.65 & 17.32 & 17.30 & 17.11 & 17.21 &  $-$0.21 &  0.13  \\
147  & SDSSp J153612.97+002039.5 & \othername & 19.26 & 19.17 & 19.45 & 19.34 & 19.26 &   0.17 &  0.04  \\
148  & SDSSp J153938.01+011124.2 & \othername & 16.94 & 16.78 & 17.59 & 17.33 & 17.05 &   0.55 &  0.07  \\
\enddata

\tablenotetext{}{Positions are in J2000.0 coordinates; asinh magnitudes (\cite{LGS99}) are
quoted. For reference, zero flux corresponds to asinh magnitudes of 
23.40, 24.22, 23.98, 23.51, and 21.83 in $u^*, g^*, r^*, i^*$, and $z^*$,
respectively. Photometric errors are typically 0.03$^m$ (see \S 2). Astrometric errors 
are typically 0.1 arcsec.}
\tablenotetext{a}{Names from the Henden \& Stone (1998) list (except for K14090198 which they
discovered later from additional data).}
\tablenotetext{b}{The mean $r^*$ magnitude determined from two measurements.}
\tablenotetext{c}{The mean  $g^*-r^*$ color determined from the mean $g^*$ and 
                  $r^*$ magnitudes.}

\end{deluxetable}
\end{scriptsize}


\newpage
\begin{figure}
\plotfiddle{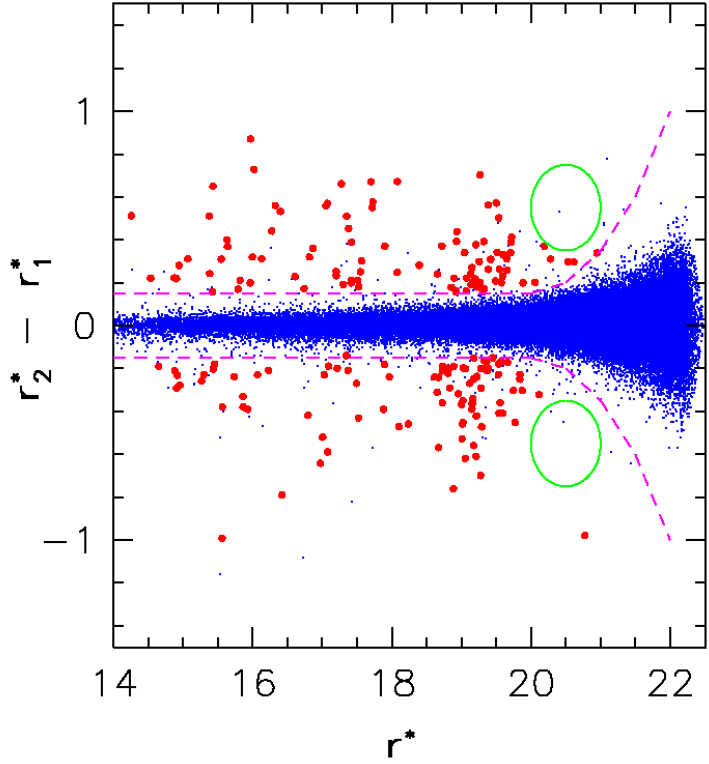}{8cm}{0}{50}{50}{-185}{-30}
\caption{
a) Observed change of $r^*$ magnitude plotted
against the mean $r^*$ magnitude for 90569 unresolved sources with
$-0.1 < g^* - r^* < 0.4$, marked as small dots. Large dots
mark 186 sources  which satisfy 
$|\Delta g^*| > 0.15^m$, 
$|\Delta r^*| > 0.15^m$,  
at least 5 $\sigma$ variability in both g' and r' bands, and 
which are brighter in $r^*$ when they are bluer in $g^* - r^*$. 
The two dashed lines show the boundary of the variability cutoff. 
The lack of faint objects with large amplitudes in the regions outlined 
by the two ellipses indicates that we are detecting the faint 
end of the RR Lyrae magnitude distribution and hence the limit 
of their distance distribution. Note that the absence of such 
sources is not due to our selection criteria, since sources in 
those two regions are already absent in the starting sample
(dots). 
}
\end{figure}

\setcounter{figure}{0}
\begin{figure}
\plotfiddle{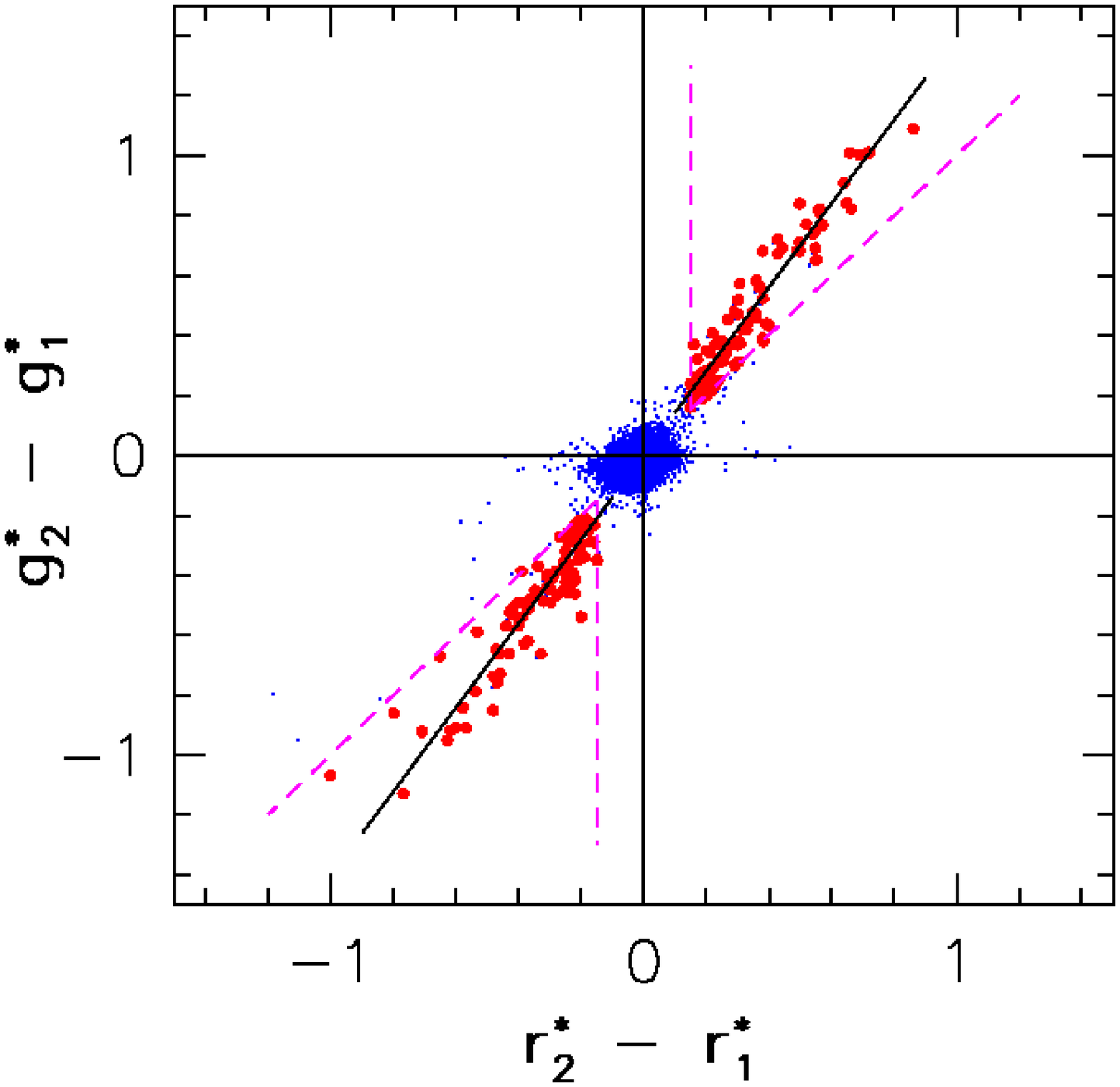}{8cm}{0}{50}{50}{-185}{-30}
\caption{
b) Observed change of $g^*$ magnitude plotted
against the observed change of $r^*$ magnitude for
sources from Figure 1a which satisfy 
$r^* < 20$. The meaning of the symbols is the same as in Figure 1a.
Vertical dashed lines show the selection conditions 
$|\Delta g^*| > 0.15^m$, $|\Delta r^*| > 0.15^m$, and 
the diagonal dashed line shows the condition 
$|\Delta g^*| > |\Delta r^*|$ (or equivalently, brighter in $r^*$
when bluer in $g^* - r^*$). The diagonal solid line shows a 
best-fit relation $|g^*_2-g^*_1| = 1.4 \, |r^*_2-r^*_1|$. 
}    
\end{figure}

\begin{figure}
\plotfiddle{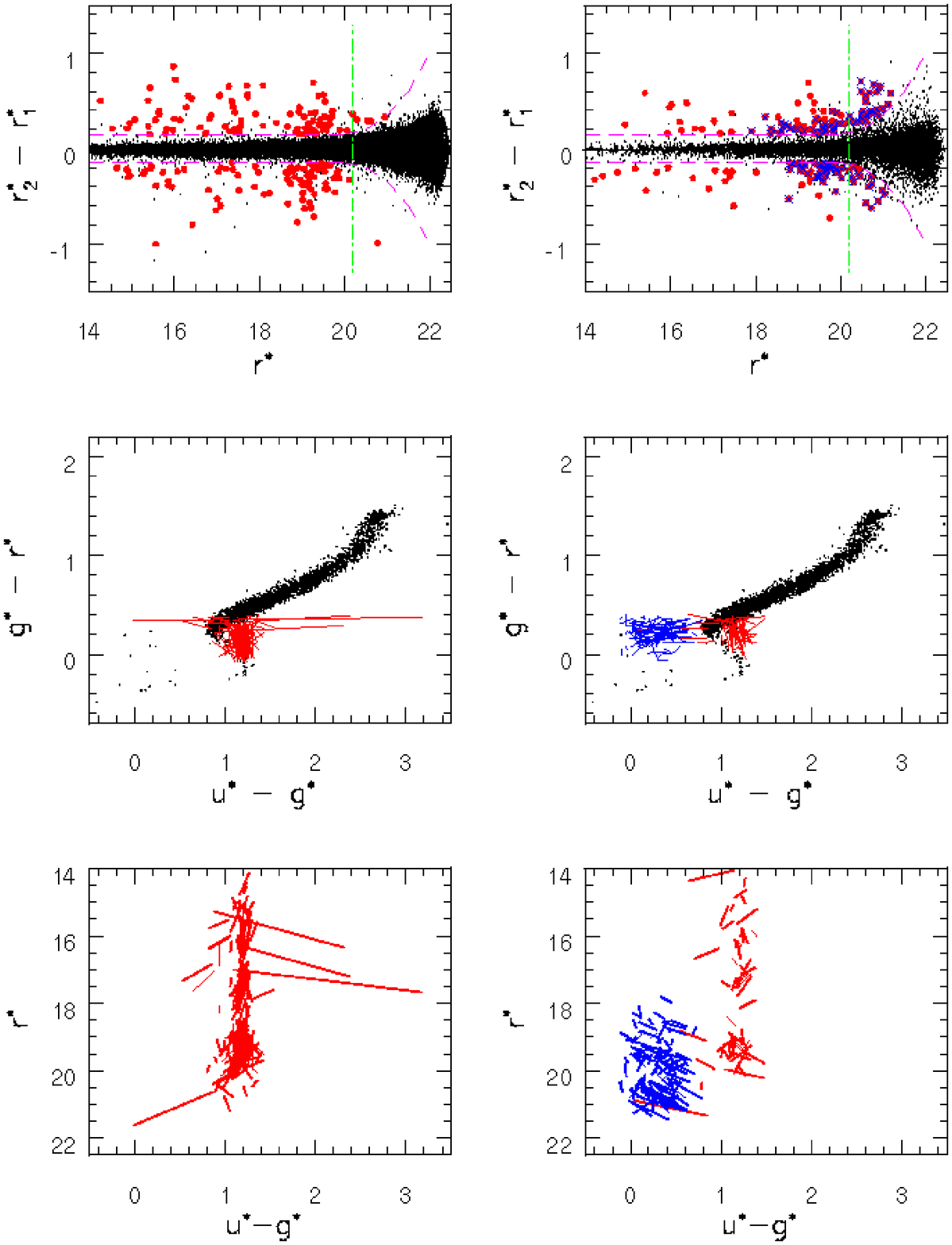}{8cm}{0}{40}{40}{-185}{0}
\caption{Comparison of two searches for RR Lyrae stars: the left
column is for 100 deg$^2$ of sky observed 2 days apart, and the 
right column is for 35 deg$^2$ of sky observed 9 months apart. 
The top two panels are analogous to Figure 1a (the distribution
of nonvariable objects in the left panel appears to be somewhat
wider because it includes $\sim$ 3 times more sources). Vertical 
lines at $r^* = 20.2$ are added to guide the eye and mark the faint 
end of the RR Lyrae star magnitude distribution. Sources with 
$u^*-g^* < 0.8$ are marked by crosses and represent variable 
QSOs rather than RR Lyrae stars. Their color difference can be 
seen in the two middle panels, which show $u^* - g^*$ vs. 
$g^* - r^*$ diagrams. In these diagrams, variable sources are 
marked by lines which connect photometric measurements at the 
two epochs. Dots represent a subsample of 5000 nonvariable objects 
and clearly outline the stellar locus. Note that there are no variable 
sources with $u^*-g^* < 0.8$ in the left panel since QSOs do not 
vary sufficiently on a two-day timescale. The lower two panels display 
$r^*$ vs. $g^* - r^*$ color-magnitude diagrams, and show that we detect no 
RR Lyrae stars fainter than $r^* \sim 20$ even though variable QSOs, 
which are selected by identical criteria, are detected to $r^* > 21$.}
\label{compare}
\end{figure}

\begin{figure}
\plotfiddle{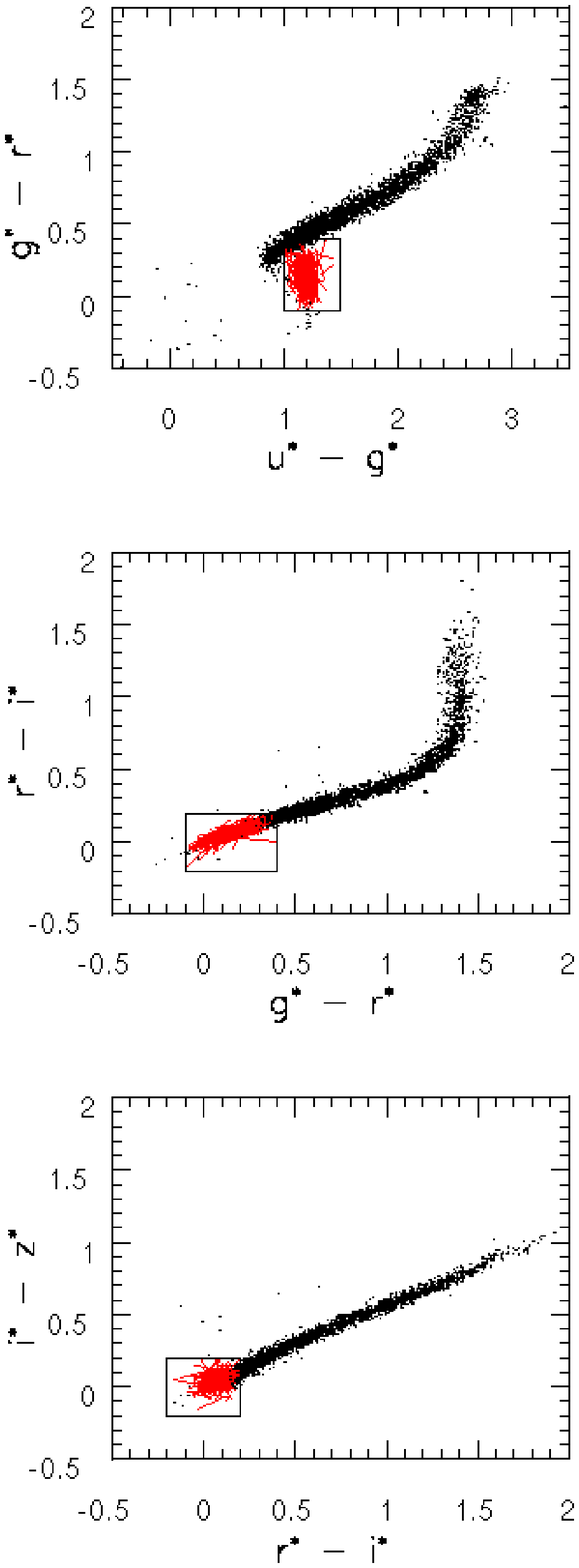}{8cm}{0}{40}{40}{-185}{0}
\caption{Color-color diagrams displaying the final RR Lyrae sample 
of 148 stars marked by lines which connect photometric measurements
at the two epochs. Dots represent a subsample of nonvariable unresolved 
objects and outline the stellar locus. The final RR Lyrae sample was
selected from the 186 candidates shown in Figure 1a by imposing color cuts 
displayed as boxes.}
\label{ccdiags}
\end{figure}

\begin{figure}
\plotfiddle{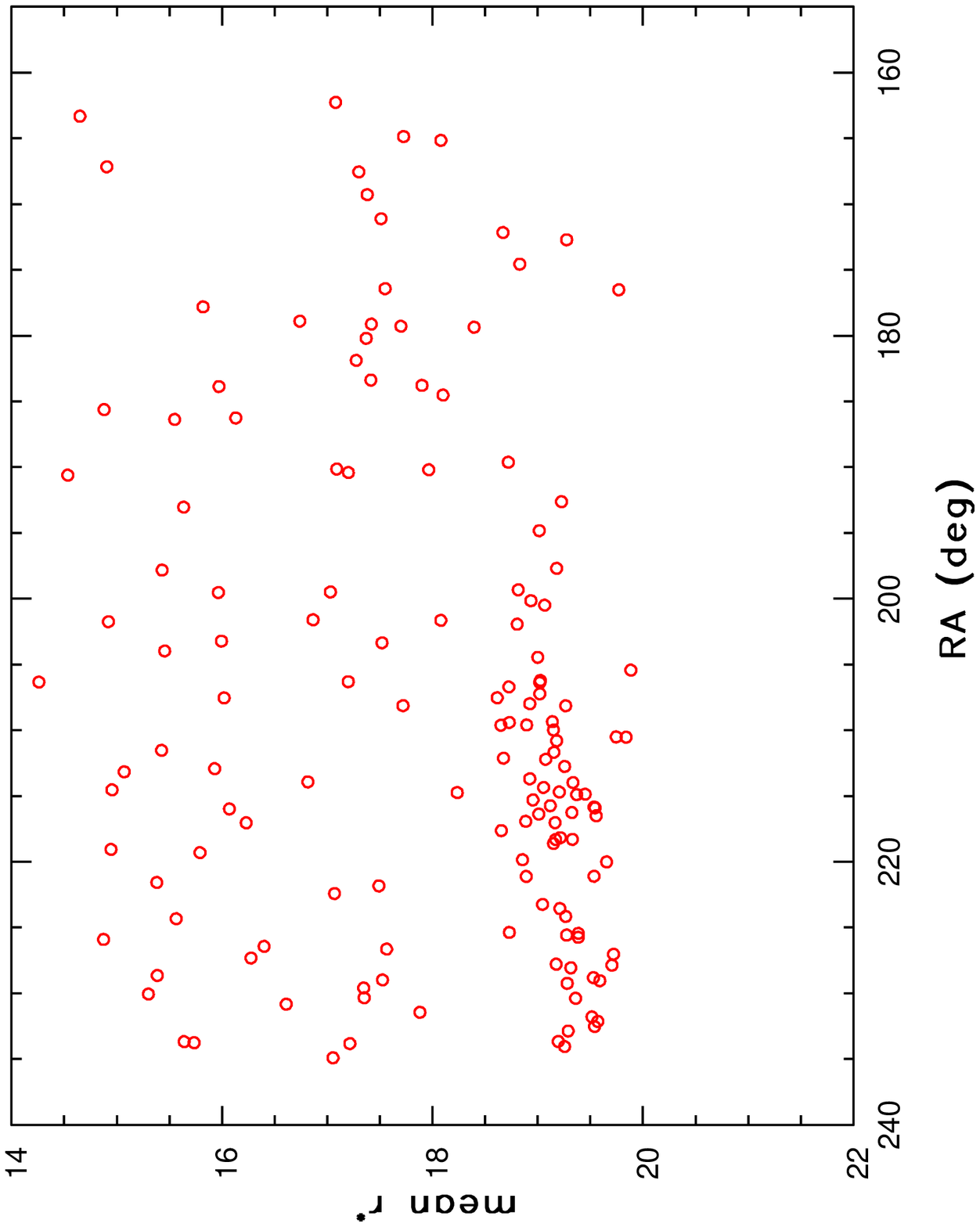}{10cm}{-90}{60}{60}{-185}{350}
\caption{Distribution of \RRc, marked as open circles,
in the mean $r^*$ vs. RA diagram. Note the concentration of sources 
with $r^* \sim 19-19.5$ and 205$^\circ$ $<$ RA $<$ 230$^\circ$.} 
\label{rRA_RR}
\end{figure}

\begin{figure}
\plotfiddle{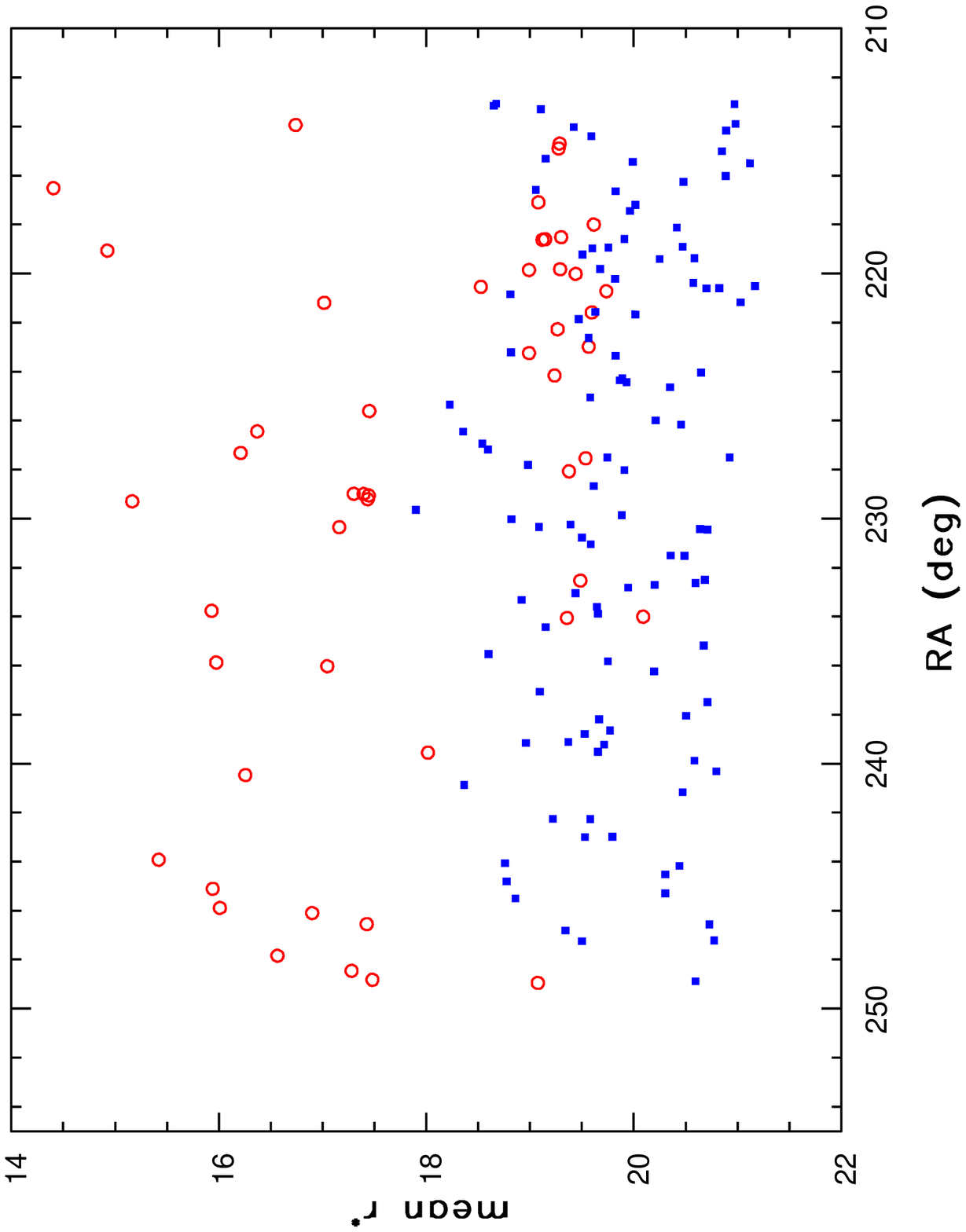}{10cm}{-90}{60}{60}{-185}{350}
\caption{Distribution of \RRc\ detected in two runs 
obtained 9 months apart in the mean $r^*$ vs. RA diagram. Sources
with $u^*-g^* > 0.8$, presumably RR Lyrae stars, are marked by open 
circles and the sources with $u^*-g^* < 0.8$, presumably QSOs, are 
marked by solid squares. Note that the distribution of RR Lyrae
stars, most notably the concentration of sources with 
$r^* \sim 19-19.5$ at 215$^\circ$ $<$ RA $<$ 230$^\circ$, is markedly 
different from the homogeneous distribution of QSOs, even though both are 
selected by identical criteria.} 
\label{rRA_A}
\end{figure}

\begin{figure}
\plotfiddle{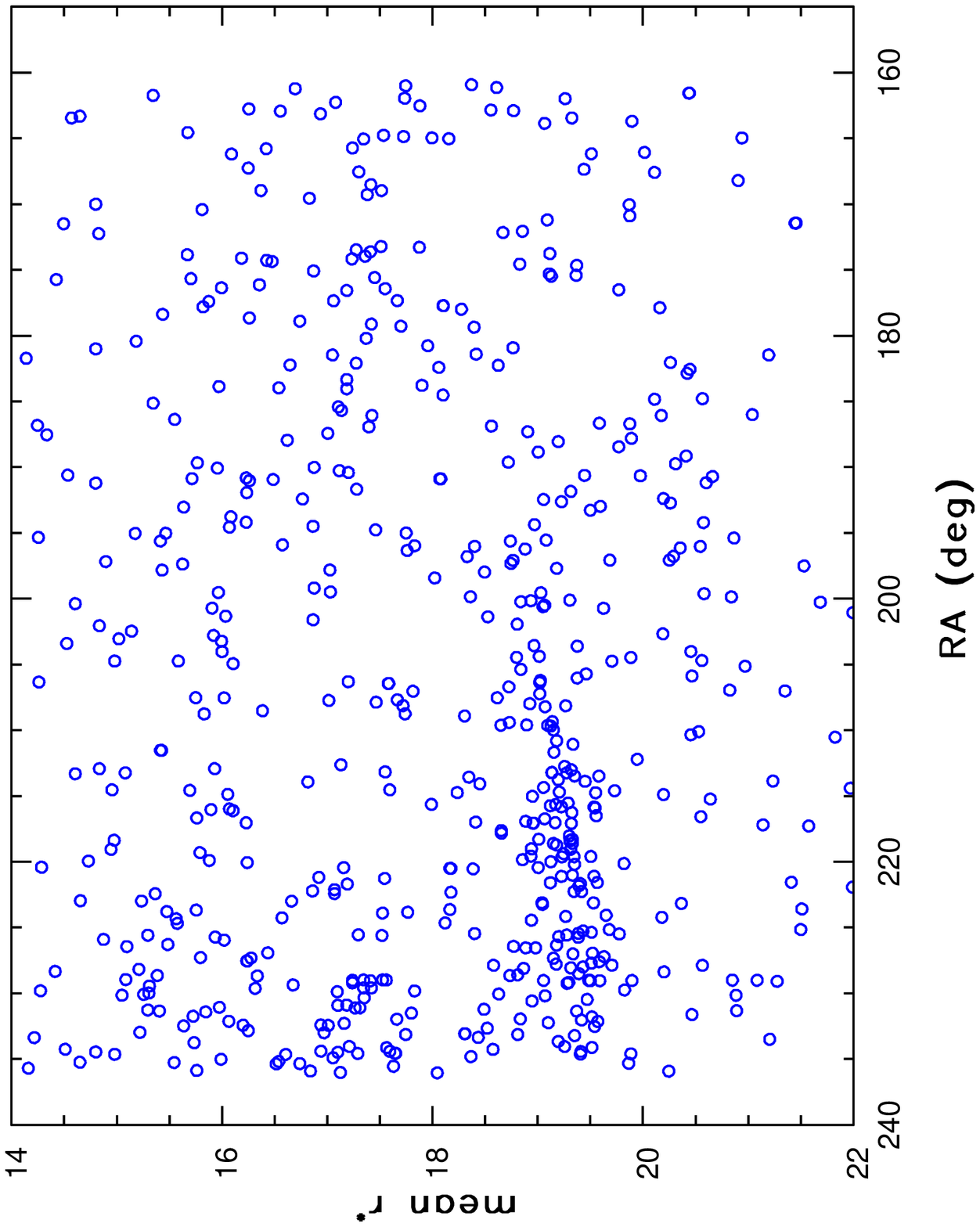}{10cm}{-90}{60}{60}{-185}{350}
\caption{Distribution of 587 stars from the overlap of runs 745 and
756 satisfying $1.1 < u^*-g^* < 1.5$ and $-0.1 < g^*-r^* < 0.3$ 
in the mean $r^*$ vs. RA diagram. Note the concentration of 
sources with $r^* \sim 19-19.5$ and 205$^\circ$ $<$ RA $<$ 230$^\circ$, 
the same magnitude--RA range as for the concentration of \RRc\
displayed in Figure 4.}
\label{rRA_9months}
\end{figure}

\begin{figure}
\plotfiddle{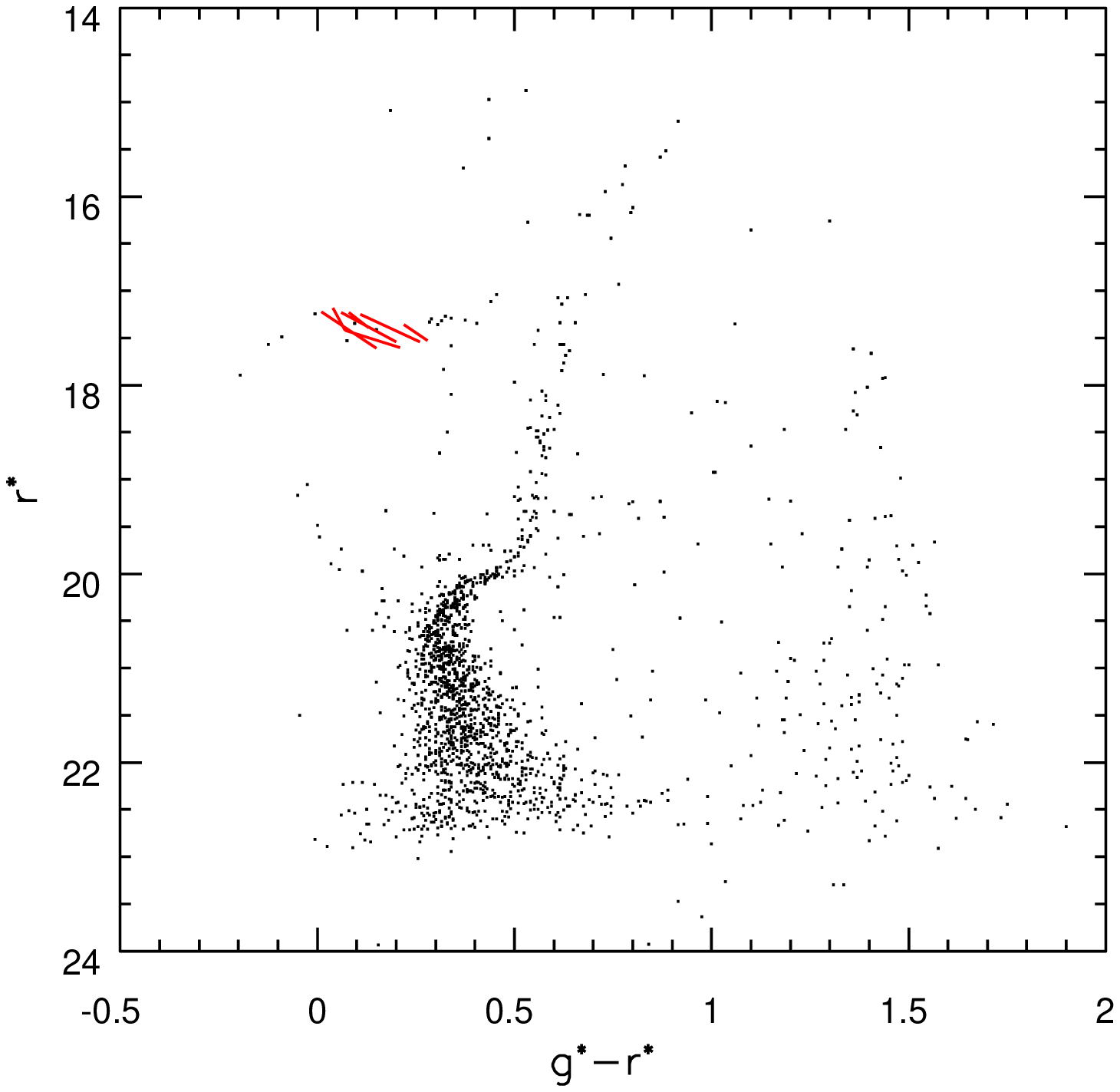}{10cm}{0}{60}{60}{-185}{0}
\caption{$r^*$ vs. $g^*-r^*$ color-magnitude diagram for $\sim 2000$ stars 
observed in SDSS commissioning run 756 inside a circle with 5 arcmin 
radius and centered on the position of the core of globular cluster Palomar 5.
Five stars selected as \RRc\ are marked by lines which connect measurements 
at different epochs; all fall in the appropriate blue horizontal branch 
region for Palomar 5 ($g^* \sim$ 17.5).}
\label{Pal5CMD}
\end{figure}

\begin{figure}
\plotfiddle{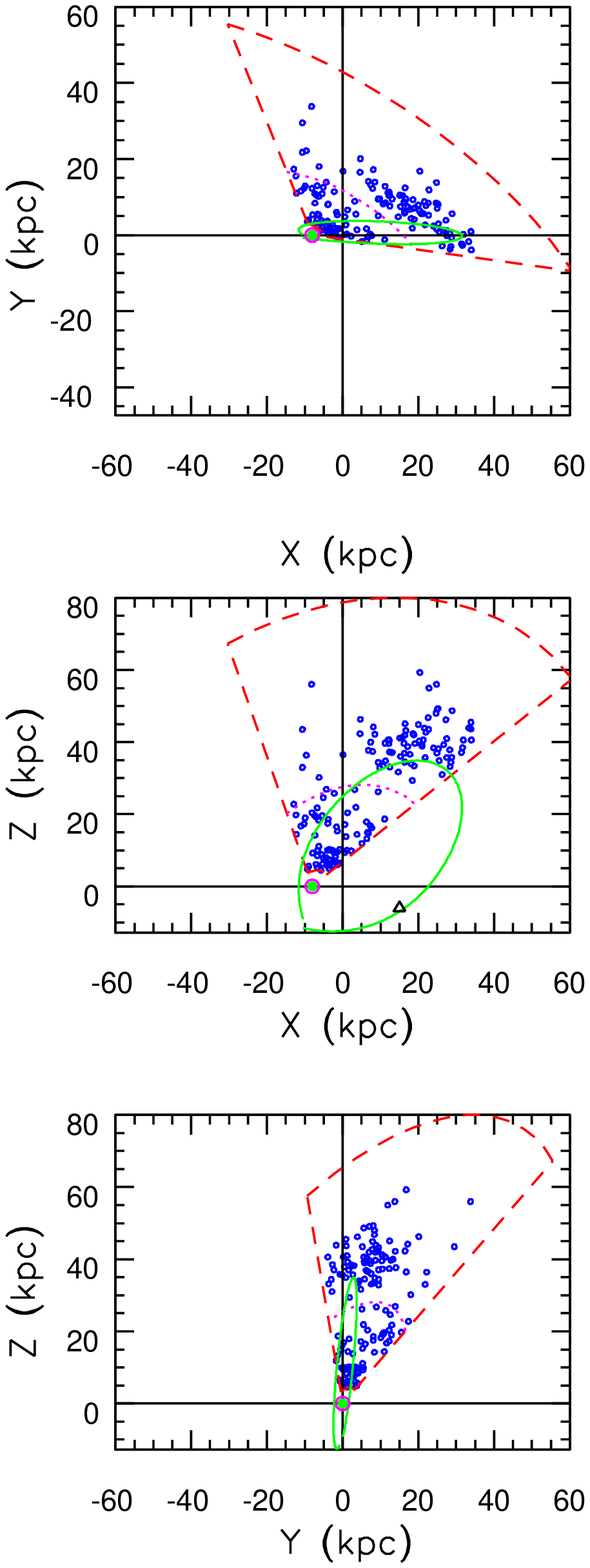}{8cm}{0}{50}{50}{-185}{-30}
\caption{Galactic distribution of selected \RRc, 
displayed as small open circles. The Sun (marked by a large dot) is at 
(X=$-8$ kpc, 0, 0). The dashed lines
show the volume within which our data can detect RR Lyrae stars: a very 
thin wedge with an opening angle of 80$^\circ$ and
distances ranging from 5 kpc (saturation limit) to 90 kpc (faint limit).
Dotted lines are added to guide the eye and show the intersection of 
this wedge with a Galactocentric sphere of radius 30 kpc ($r^* \sim$ 18-18.5). 
Note the group of \RRc\ at (X=20 kpc, Y=10 kpc, Z=40kpc), corresponding 
to the concentration of sources with $r^* \sim 19-19.5$ and 
205$^\circ$ $<$ RA $<$ 230$^\circ$ visible in Figure 4. The solid line displays the
orbit of the Sgr dwarf spheroidal (situated at X=15 kpc, Y=$-2$ kpc, Z=$-6$
kpc and marked by a triangle in the middle panel) as calculated by Johnston 
{\em et al.} (1999a).}
\label{galaxy}
\end{figure}

\begin{figure}
\plotfiddle{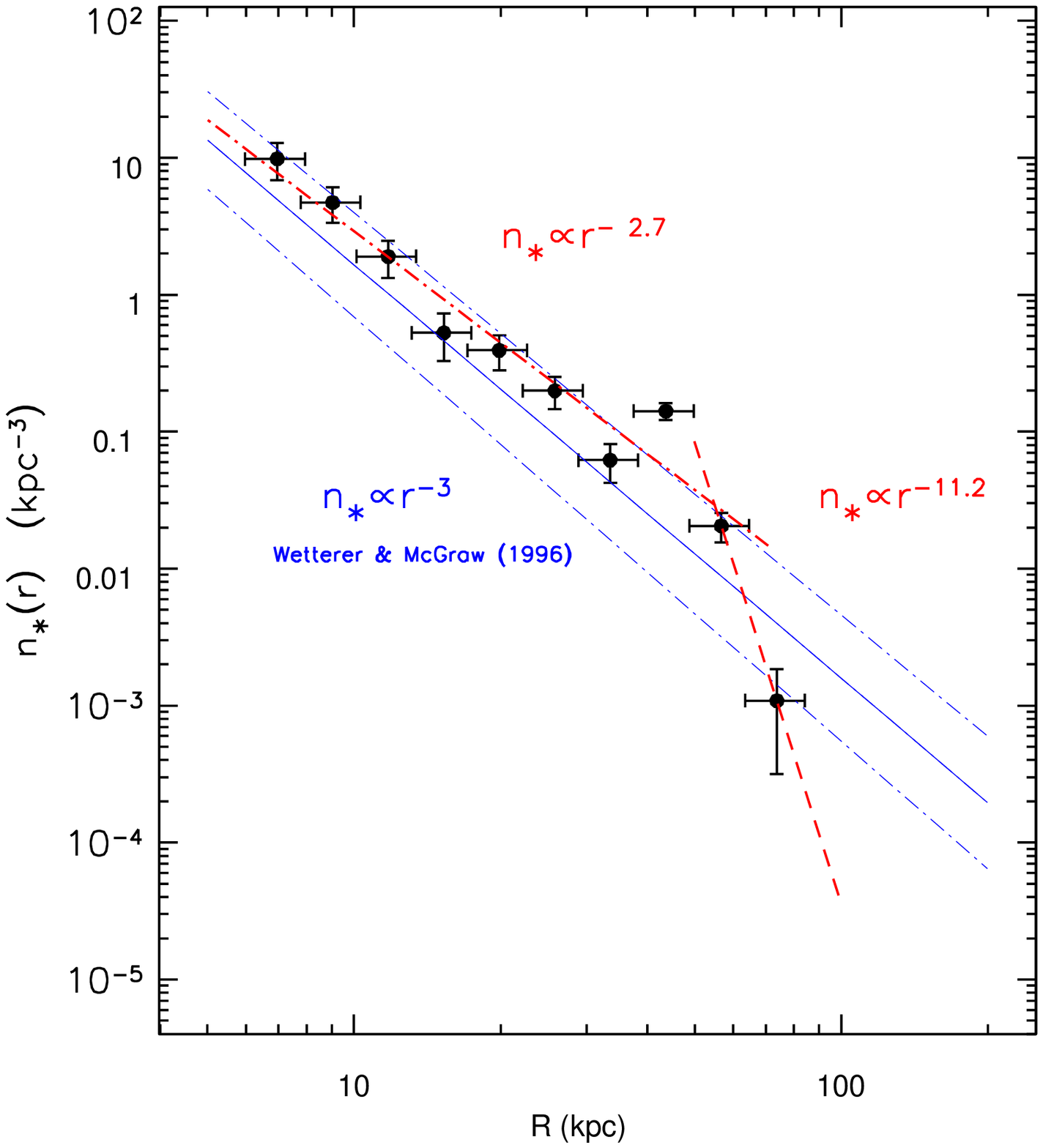}{10cm}{0}{60}{60}{-185}{0}
\caption{Comparison between the volume density for RR Lyrae stars 
obtained from our data (dots with error bars) and the $R^{-3}$ power 
law determined by Wetterer \& McGraw (1996, thick solid line with the 
1$\sigma$ uncertainty shown by thin dot-dashed lines). We find that the 
volume density follows a shallower power law with a best-fit power 
index of 2.7 $\pm$ 0.2 (thick dot-dashed line) for $R <$ 50 kpc, and 
turns off sharply as a rather steep $R^{-11.2}$ power law for $R >$ 
60 kpc (thick dashed line). Data for $R <$ 35 kpc are consistent with
the $R^{-3}$ distribution.}
\label{nofR}
\end{figure}

\begin{figure}
\plotfiddle{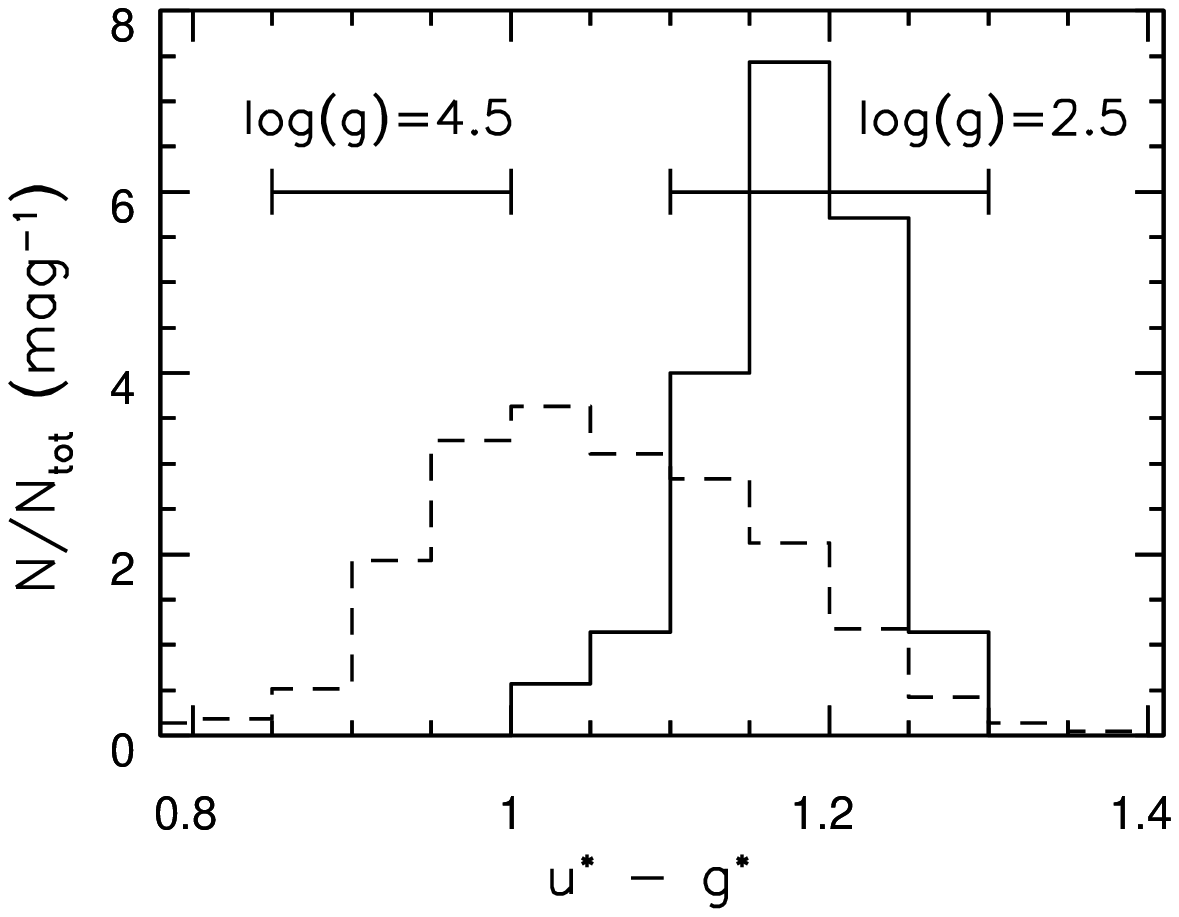}{10cm}{0}{60}{60}{-185}{0}
\caption{The $u^*-g^*$ color distribution for stars selected by 
$0.0 < g^*-r^* < 0.1$ (c.f. Figure 2) from the 90569 stars shown in 
Figure 1a (dashed line) and for stars selected by the same criterion 
from the resulting sample of candidate RR Lyrae stars (solid line). 
Note that the $u^*-g^*$ color of candidate RR Lyrae stars is on average 
redder for $\about 0.2$ mag than the $u^*-g^*$ color of nonvariable stars 
within the same narrow range of $g^*-r^*$ color. Theoretical expectations 
taken from Lenz et al. (1998) are shown as horizontal lines. For Main 
Sequence stars the expected $u^*-g^*$ range is 0.85--1.0 (marked as log(g)=4.5) 
and for Horizontal Branch stars the expected $u^*-g^*$ range is 1.1--1.3 
(marked as log(g)=2.5). The intrinsic spread of the $u^*-g^*$ color for a given 
gravity is due to varying metalicity and the plotted values correspond 
to the range $-2.0 < [M/H] < 0$.}
\label{ughist}
\end{figure}

\begin{figure}
\plotfiddle{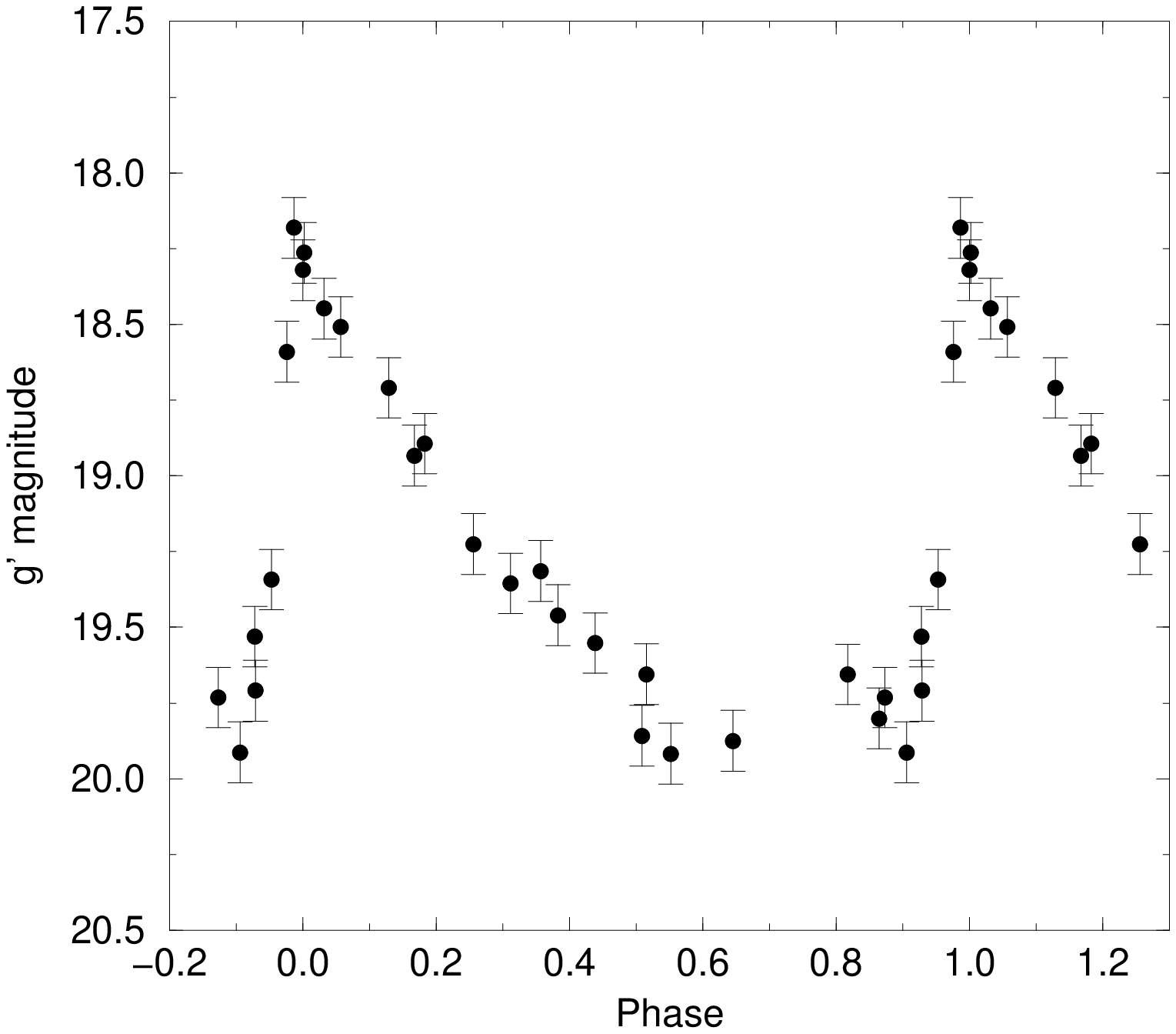}{10cm}{0}{60}{60}{-185}{0}
\caption{
a) Light curve in $g'$ band for SDSSp J113049.26$-$005918.2 obtained by the SDSS
photometric telescope. The shape of the light curve and its period (0.46379 d) 
confirm that this candidate is an RR Lyrae star.  
}
\end{figure}

\setcounter{figure}{10}
\begin{figure}
\plotfiddle{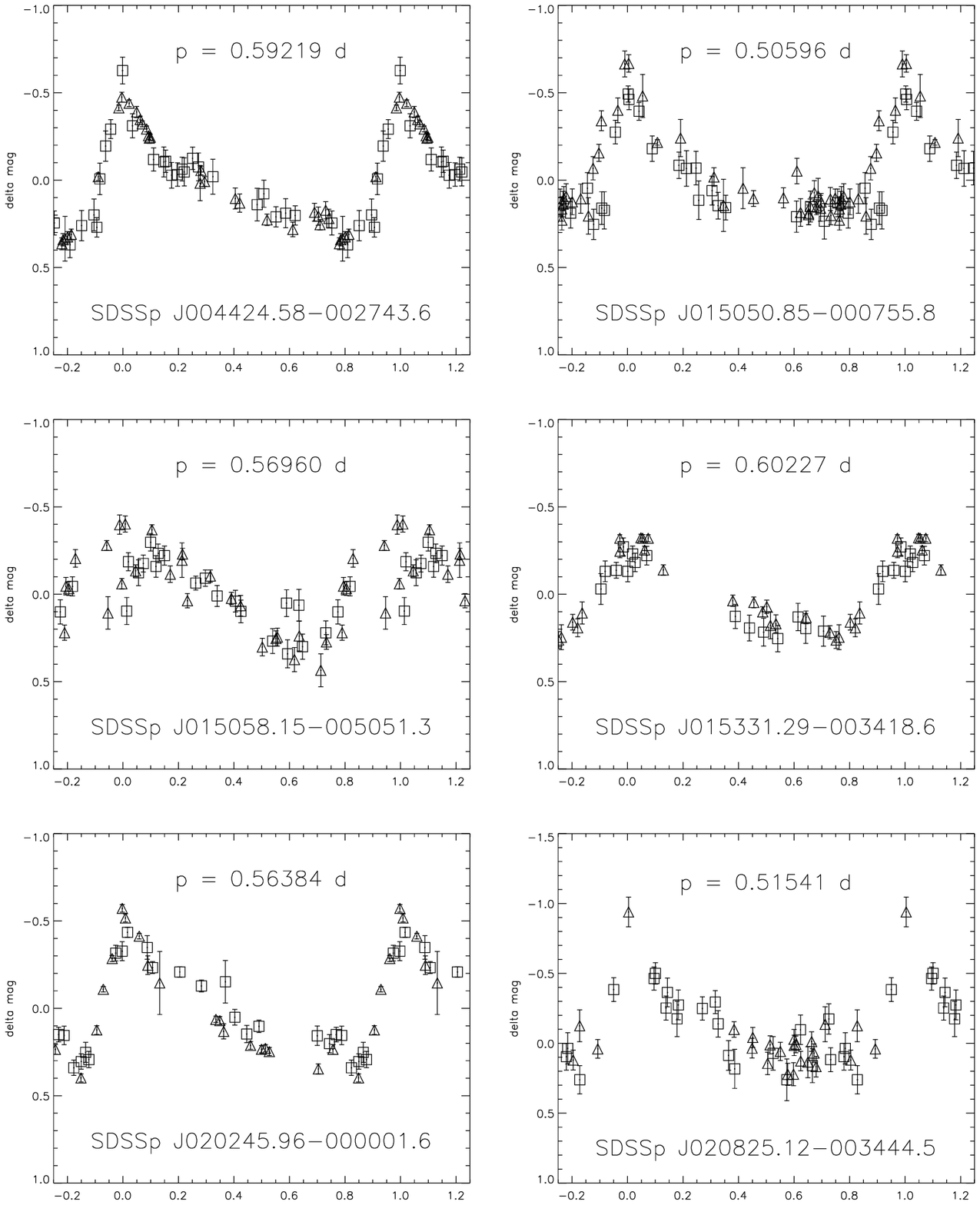}{8cm}{0}{60}{60}{-185}{-30}
\caption{
b) Johnson V band light curves for 6 candidate RR Lyrae stars (not listed in Table 2) 
selected from SDSS commissioning data obtained in the southern Galactic 
hemisphere by the method described here. The light-curve data was obtained by 
the 0.76-m reflector of the University of Washington's Manastash Ridge
Observatory (triangles) and supplemented by the data from the LONEOS
database (squares). All 6 candidates have light curves with shapes and periods 
characteristic of RR Lyrae stars.
}
\label{lightcurves}
\end{figure}

\end{document}